\documentclass{aa}  

\usepackage{graphicx}
\usepackage{txfonts}
\usepackage{natbib}
\usepackage[citecolor=blue,colorlinks=true,linkcolor=blue,urlcolor=blue]{hyperref}
\graphicspath{{./}{figures/}}
\usepackage[export]{adjustbox}
\usepackage{tablefootnote}

\begin{document}

   \title{G189.6+03.3: the first complete X-ray view provided by SRG/eROSITA}


\author{Francesco Camilloni 
    \inst{1}\thanks{\email{fcam@mpe.mpg.de}}
    \and Werner Becker \inst{1}\inst{,2}
}

\institute{Max-Planck-Institut für extraterrestrische Physik, Giessenbachstraße, 85748 Garching, Germany    
 \and
Max-Planck-Institut für Radioastronomie, Auf dem Hügel 69, 53121 Bonn, Germany
}
   \date{Received 27 June 2023 / Accepted 28 September 2023}

 
  \abstract{G189.6+03.3 and IC443 are two examples of supernova remnants located in a region rich of gas and dust, spatially close to the HII region S249. So far, the actual shape of IC443 is believed to be given by the past action of multiple supernova explosions, while a third unrelated might have originated G189.6+03.3.}
  {If the IC443 nebula has been extensively observed in several bands, in opposite there is an almost complete lack of observations on the nearby and much weaker supernova remnant G189.6+03.3, discovered in 1994 with \textit{ROSAT}. Given the relatively large extent of this second remnant, the new dataset provided by the X-ray telescope eROSITA onboard the Spectrum Roentgen Gamma (\textit{SRG}) mission gives a unique opportunity to characterize it more in depth.}
  {We provide a full spectral characterization of G189.6+03.3 emission for the first time, together with new images covering the whole remnant. Since one of the leading hypothesis is that its emission partially overlaps with the emission of IC443, we test this scenario dividing the remnant in several regions from which we extracted the spectra.}
  {The new X-ray images provided by eROSITA show an elongated structure. Together with the detection of supersolar abundances of O, Mg, Ne and Si and subsolar abundance of Fe, these features could be an indication of a faint supernova explosion. The X-ray spectra also highlight the presence of a 0.7 keV plasma component across all the regions together with a column density almost uniform. }
  {The ubiquitous presence of the 0.7 keV plasma component is a strong indication for G189.6+03.3 overlapping completely with IC443. We propose the progenitors of G189.6+03.3 and IC443 could have been hosted in a binary or multiple system, originating two explosions at different times in different positions.}
\authorrunning{F.Camilloni \& W. Becker}
\titlerunning{G189.6+03.3: the first X-ray view provided by SRG/eROSITA}

   \maketitle

   %
\section{Introduction \label{sec:intro}}
The majority of the known supernova remnants (SNRs) have been discovered in and identified by their radio band. That a supernova remnant has to be radio bright was therefore considered to be a necessary condition before a
diffuse and extended source, e.g. detected at X-ray energies, was generally accepted as being an SNR. An example of such a source is G189.6+03.3 which has remained largely uncharted until today. It was clearly identified as an SNR by \cite{Asaoka1994} using data from the \textit{ROSAT} observatory obtained during the first ever imaging X-ray All-Sky Survey (RASS). However, the lack of its radio counterpart for quite some years lead to an SNR candidate status which still holds on in the literature till today, regardless that its radio emission was detected by \cite{Leahy2004} some years after its discovery.  

Following studies on the much brighter supernova remnant G189.1+03.0 (IC443), located at the western edge of G189.6+03.3, often did not even mention the latter. The most likely explanation for this is probably its very low surface brightness: indeed, its shape is barely visible in the early images shown by \cite{Asaoka1994}, which for many years were the only ones available from the remnant. The remnant has an extent of about 0.75 degrees radius. The distance estimated for G189.6+03.3 is 1.5 kpc, while its age estimate is $3\times 10^{4}$ years \citep{Asaoka1994}.

Nevertheless, \cite{Asaoka1994} provided a comprehensive overview of the physical properties of G189.6+03.3 and a clear detection at radio wavelength was provided by \cite{Leahy2004} some years after its discovery. Although the sensitivity and spectral resolution of \textit{ROSAT} was modest when compared with more recent observatories, the authors managed to estimate the mean plasma temperature of G189.6+03.3 to be at the order of 0.14 keV, with a column density between $0.6-1.3 \times 10^{22}$ cm$^{-2}$ (90\% confidence level). They also were the first who proposed that an optical filamentary structure located north of G189.6+03.3 might trace the interaction of the remnant with the HII emitting region S249 \citep{Fesen1984,Braun1986}. In addition, from the spatial distribution of the column density in some selected regions, \cite{Asaoka1994} suggest that G189.6+03.3 is overlapping with nearly half of IC443, interpreting the dark lane characterizing the images of IC443 as a result of the overlap of cold material, possibly associated to G189.6+03.3. This interpretation was suggested by IC443 being located in a very gas rich region \citep{Fesen1980,Fesen1984,Braun1986} with its progenitor probably belonging to a group of massive stars called Gem OB1 association \citep{Humphreys1978}. While \cite{Denoyer1978}, \cite{Troja2006}, \cite{Troja2008} demonstrated how IC443 is interacting with an atomic cloud, other studies \citep{Cornett1977,Burton1988,Claussen1997} have shown that also a molecular cloud is interacting with IC443. This confirms that the environment surrounding the remnant is very rich of gas. According to \cite{Braun1986}, strong winds and X-ray emission from these massive stars probably carved a system of cavities where at least one massive star exploded, forming IC443. Therefore, it is not unlikely that another massive star belonging to this association formed G189.6+03.3 well before IC443 was formed. 

The only other X-ray observation of a part of G189.6+03.3 has been obtained with \textit{Suzaku} \citep{Mitsuda2007} on a bright knot located near the north-eastern part of the remnant, almost on the opposite side of where the spectral analysis of \cite{Asaoka1994} was performed. The findings of \cite{Yamauchi2020} are particularly important because they find evidence of Radiative Recombination Continuum (RRC) emission around 2.5 keV. This demonstrated the presence of a recombining plasma, similarly of what was discovered by \cite{Yamaguchi2009} for the nearby SNR IC443. 

 Therefore, the launch of the eROSITA telescope in July 2019  \citep{Predehl2021} onboard the Spectrum Roentgen Gamma (\textit{SRG}) mission \citep{Sunyaev2021} provided a new opportunity to study extended sources like G189.6+03.3 due to the telescope's unlimited field of view (FOV) in its all-sky survey mode. At the date of February 2022, eROSITA has completed about four and a third all-sky surveys. The instrument is made by seven independent telescope modules (TMs), providing a large effective area \cite[see for details][]{Predehl2021}. Moreover, the almost $\sim1^{\circ}$ field view and a spectral resolution superior to that of the EPIC-PN camera onboard \textit{XMM-Newton}, coupled with the all-sky mode, makes this instrument unique for the study of SNRs and many other extended sources. We therefore employed the dataset provided by eROSITA to perform the first spatial and spectral analysis of G189.6+03.3 in its full extent. 
 
 We divide the paper as follows: in Section \ref{sec:Data Reduction} we describe the reduction and processing of the data, in Sections \ref{sec:Spatial Analysis} and \ref{sec:spectral analysis} we present the results obtained by analyzing the images and the spectra of G189.6+03.3. In Section \ref{sec:open_clusters} we describe the diffuse emission observed around two nearby star clusters, M35 and NGC 2175. In Section \ref{sec:Discussion} we discuss our results in the light of the current supernova explosion models in order to pinpoint the relation between G189.6+03.3 and IC443 finally.

\section{Data Reduction \label{sec:Data Reduction}}
The location of G189.6+03.3 was in the eROSITA field of view during all four sky surveys, i.e. between 2020 April 1st - 8th, October 3rd - 12th, 2021 March 29th - April 7th as well as  September 29th - October 9th. These observations yield an unvignetted exposure time for G189.6+03.3 of 830s when including all seven telescope modules. In contrast, the deadtime and vignetting corrected observing time is computed to be slightly less than half of that, i.e. 390s only. To perform the data reduction, the creation of images and the extraction of energy spectra, we employed the eROSITA scientific analysis software (eSASS) version 211214 and the instrument calibration files included in this software \citep{Brunner2022}\footnote{The software version is denoted according to the date when it was released, i.e. 14.12.2021.}. Within the eSASS pipeline, X-ray data of the eRASS sky are divided into 4700 partly overlapping sky tiles of $3\fdg6 \times 3\fdg6$ each. These are numbered using six digits, three for RA and three for Dec, representing the sky tile center position in degrees.
Employing the command \texttt{evtool}, we started merging 
the sky tiles numbered 092066, 094069 and 096066 from the eRASS:4 dataset (events from all four scans) to produce a merged single event file. The eSASS command \texttt{radec2xy} was applied to align the images on the center region of G189.6+03.3, i.e. to the position  RA:06h18m30s DEC:+22d10m00s. For imaging we used the events from all seven telescope modules (TM 1-2-3-4-5-6-7) with the CCD detection PATTERN=15 and filtering on the good time intervals (absence of solar flaring). We then extracted spectrum, background, Redistribution Matrix File (RMF) and Ancillary Response Files (ARF) applying the command \texttt{srctool} on data from TM 1-2-3-4-6 only. A light leak was detected soon after the launch of eROSITA for the telescope modules TM 5 and 7 \citep{Predehl2021}. As this light leak introduces an extra uncertainty when it comes to the calibration of these detectors, data from TM 5 and 7 are considered to be less suitable for spectroscopic studies. In consequence, events from these telescope modules were not included in the spectral analysis.

\section{Spatial Analysis \label{sec:Spatial Analysis}}
\begin{figure*}
    \centering
    \includegraphics[scale=0.7]{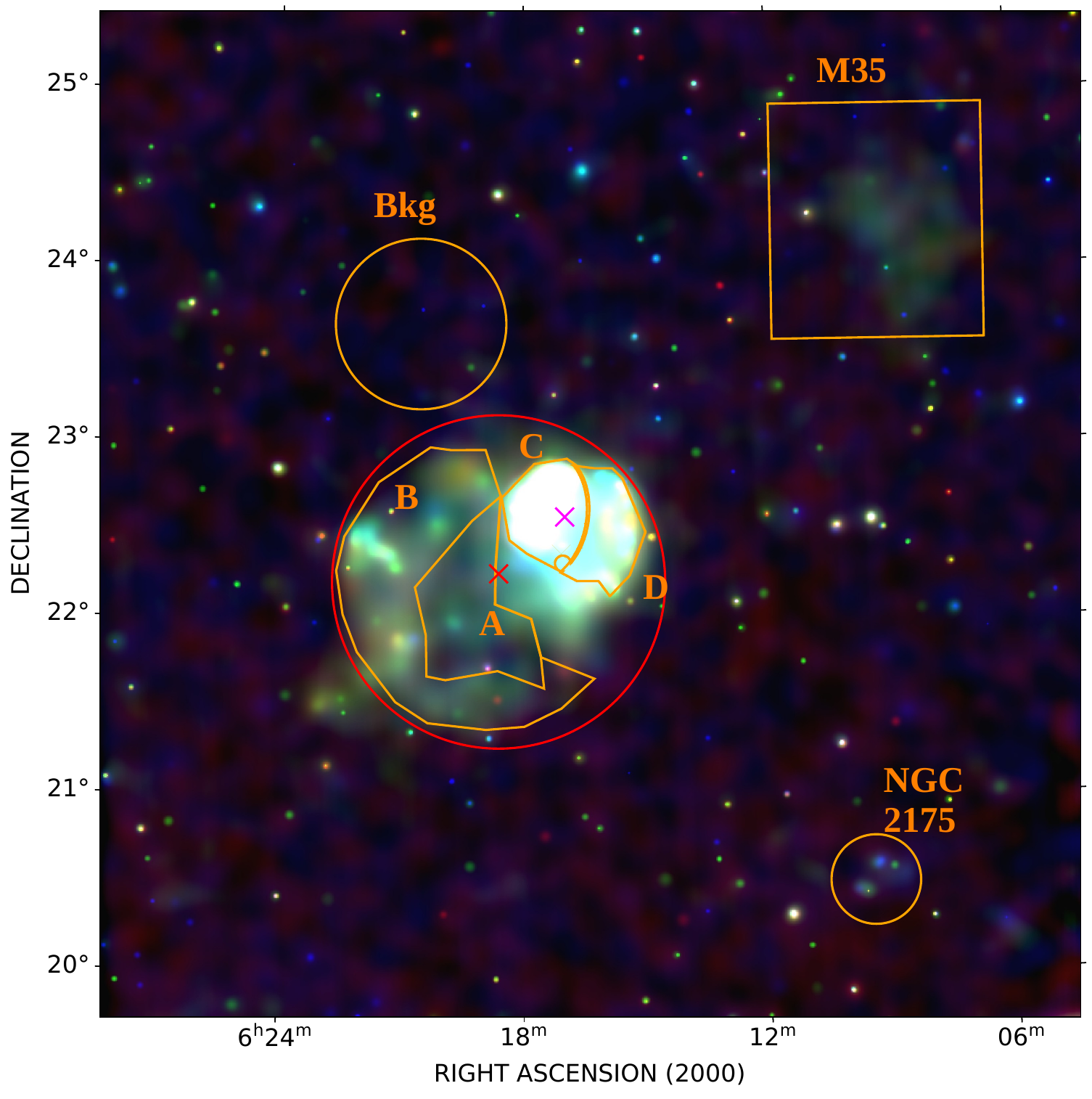}
    \caption{False color image (Red: 0.2-0.7 keV, Green: 0.7-1.1 keV, Blue: 1.1-10 keV) of the supernova remnant IC443 and supernova remnant G189.6+03.3 obtained with eRASS:4 dataset. For each of the three images, we applied the adaptive smoothing algorithm of \cite{Ebeling2006} with a  minimum significance of the signal S/N=3 and 5 as maximum. The minimum scale of smoothing is pixel size, while the maximum is 8 pixels. The scale of the colors has been particularly stretched to highlight the diffuse emission. In orange we display the extraction regions employed for the spectral analysis. The red circle is not used for spectral analysis purposes and it just indicative for the suggested extension of G189.6+03.3 (the red cross marks the center of the circle at RA:06h19m40.8s, DEC:+21:58:03). The magenta cross indicates the center of IC443 at RA:06h17m0s and DEC:+22:34:00.
    \label{fig:G189_IC443_RGB}}
\end{figure*}

As a first step, we produced an RGB image from the eRASS:4 dataset reduced as described in Section \ref{sec:Data Reduction}. We applied the adaptive smoothing algorithm of \cite{Ebeling2006} to enhance the diffuse emission of the G189.6+03.3 and IC443 complex. Comparing to the original image of \cite{Asaoka1994}, a considerable higher number of details appear in the eROSITA RGB image shown in Figure \ref{fig:G189_IC443_RGB}. The shape of the remnant from Figure \ref{fig:G189_IC443_RGB} is slightly asymmetric, appearing elongated in the South-East direction, but also, in the West part if we consider region 'D' as part of G189.6+03.3. In this case, the shape of the remnant becomes more symmetric, with 'ear-like' feature (see for example \cite{Grichener2017} and references therein for recent discussion about ear-like structure in SNR). 

\begin{figure}
    \centering
    \includegraphics[scale=0.6]{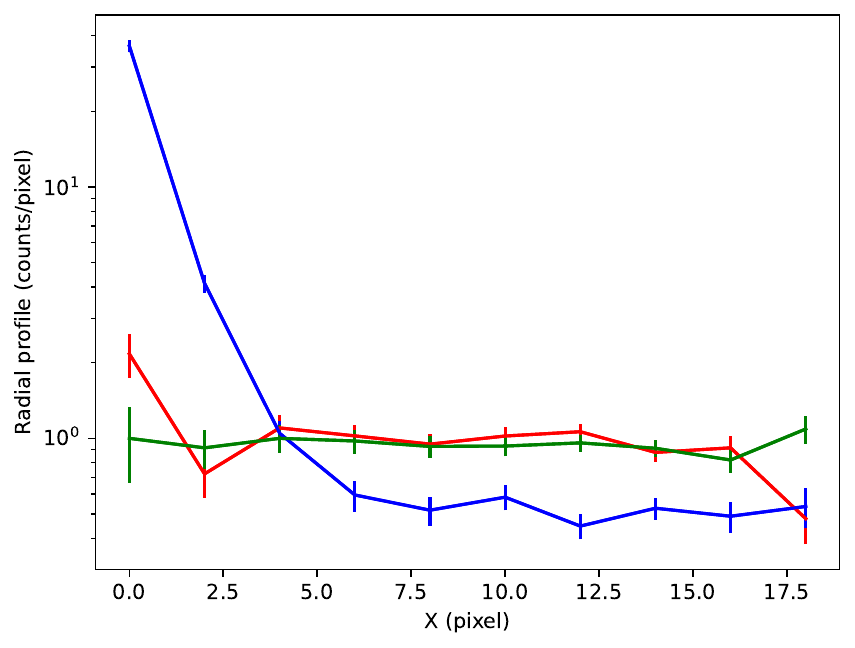}
    \caption{Radial profiles extracted respectively from the light blue circle region in Figure \ref{fig:G189_IC443_zoom} (Red), a region of the same size located inside G189.6+03.3 (Green) and another one centered on the star V398 Gem (Blue).
    \label{fig:radial_profiles}}
\end{figure}

We notice two interesting features inside the shell-like structure of G189.6+03.3. One feature is a very dim diffuse emission almost at the center, which could origin from an unresolved central source. We derived a simple test to see whether this source is extended or not, comparing its radial profile with the one of a nearby region extracted inside the remnant and the one of a known point source (nearby star V398 Gem): the result is shown in Figure \ref{fig:radial_profiles}, demonstrating this is not a point source. Nevertheless, the profile seems very similar to the one extracted for another region inside the SNR, suggesting this might just be an over density. In Section \ref{sec:spectral analysis} we describe the spectral analysis we carried out on this object.

\begin{figure*}
    \centering
    \includegraphics[scale=0.5]{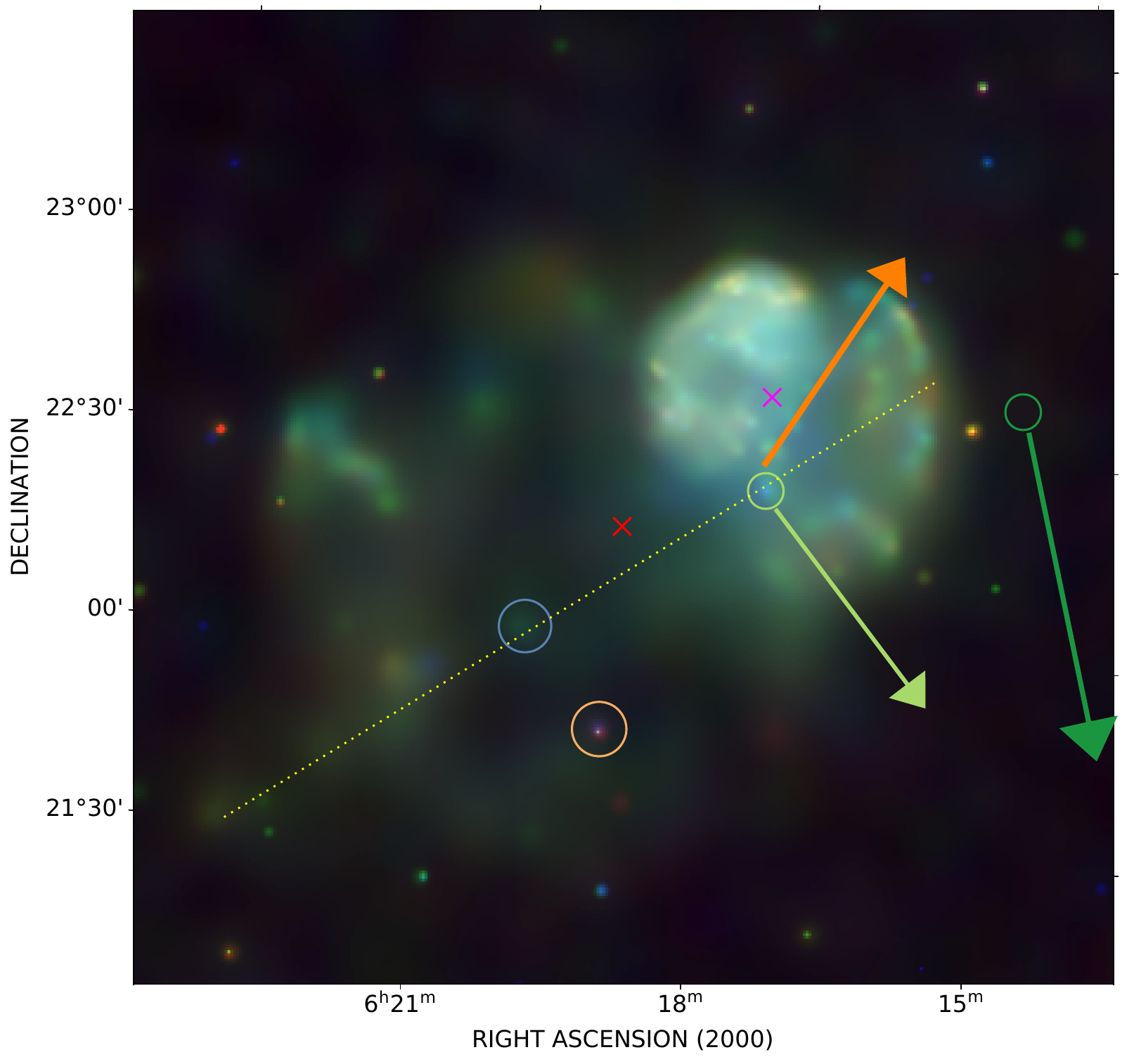}
    \caption{Close up of Figure \ref{fig:G189_IC443_RGB} zoomed on IC443 and G189.6+03.3. The image has been less stretched to show better the details within the remnant. The blue circle indicates the position of a putative central source in G189.6+03.3, in orange the transient\protect\footnotemark. The green arrow indicates the proper motion direction of the pulsar B0611+22 while the light green the one of the fast moving neutron star CXOU J061705.3+22212. Both arrows are obtained multiplying the proper motion value for the age of each one of the compact objects. For displaying purpose, the length of the arrows has been magnified 10 times. The thin dotted yellow line highlights the elongated structure visible in eROSITA, while the orange thick line represents the jet direction suggested by \cite{Greco2018}. 
    \label{fig:G189_IC443_zoom}}

\end{figure*}

A second interesting feature is an unknown bright source located at RA:6:18:53.7, DEC:+21:45:49.8 (indicated by the orange circle in Figure \ref{fig:G189_IC443_zoom}). Inspecting each single eRASS survey, it appears only in eRASS 3. Therefore, we identified the transient as described in \cite{Salvato2022} using CatWISE and Gaia EDR3 catalogs separately. The object is identified as Gaia EDR3 3376739988615000320 and AllWISE source J061853.77+214551.5 \citep{Medan2021} which is an M-type star at roughly 70 pc \citep{Zhong2019}, ruling out the possibility that it is associated with the remnant.

In the same region there are also two compact objects: a first object is the fast moving neutron star CXOU J061705.3+22212 enshrouded by a pulsar wind nebula, which was extensively observed with \textit{XMM-Newton} and \textit{Chandra} \citep{Keohane1997,Bocchino2001,Olbert2001,Gaensler2006,Swartz2015,Greco2018}. It is located at RA (J2000):06h17m05.18s and DEC (J2000):+22:21:27.6 (see Figure \ref{fig:G189_IC443_zoom}). If we assume RA (J2000):06h17m0s and DEC (J2000):+22:34:00 as center of IC443, the fast moving neutron star is 12.6 arcmin separated from it while the displacement from the center of G189.6+03.3 is 37'. CXOU J061705.3+22212 has not yet been detected as a radio pulsar. Its location was covered in the FAST GPPS survey\footnote{\url{http://zmtt.bao.ac.cn/GPPS/GPPSnewPSR.html}} which observed the source location for 18000s with the PSR backend and a limiting sensitivity of 10 $\mu$Jy \citep{Han2021}. The second nearby compact object is the radio pulsar PSR B0611+22 located at RA(J2000):06h14m17s DEC(J2000):+22:29:56.848, 1.2$^{\circ}$ from the center of G189.6+03.3 and 32' from the center of IC443. When \cite{Davies1972} discovered PSR B0611+22 in the radio band, G189.6+03.3 had not been discovered by then so that they associated the pulsar to IC443. However, today we firmly detect two compact objects and two supernova remnants spatially close to each other. 

Therefore, the immediate question is whether the two compact objects as well as IC443 and G189.6+03.3 are all at the same distance. If this turns out to be the case, we would probably be witnessing the stellar endpoint of a binary system formed by two massive stars. To test this scenario, we queried the Australia Telescope National Facility Pulsar Catalogue\footnote{\url{https://www.atnf.csiro.au/research/pulsar/psrcat/}} \citep{Manchester2005} to obtain the dispersion measure (DM) value, age, proper motion in RA (PMRA) and DEC (PMDEC) for PSR B0611+22. The dispersion measure can be correlated to the column density measured in X-rays. The dispersion measure for B0611+22 is 96.91 cm$^{-3}$ pc, and it corresponds to a distance of 1.74 kpc or 3.5 kpc, depending on the model assumed for the dispersion measure \citep[see][for a recent discussion]{Yao2017}. Assuming the relation between dispersion measure and column density given by \cite{He2013}
\begin{equation}
    \text{N}_{\rm H}(10^{20} \text{cm}^{-2}) = 0.30^{+0.13}_{-0.09}\text{ DM (pc cm}^{-3}\text{)}
\end{equation}
the equivalent value for the NH measured in X-rays is  2.9$^{+1.3}_{-0.9}\cdot10^{21}$ cm$^{-2}$. In the next Section, we will compare this value to the column density derived from the spectral fits of different regions of G189.6+03.3. 

\begin{figure}
    \centering
    \includegraphics[scale=0.35]{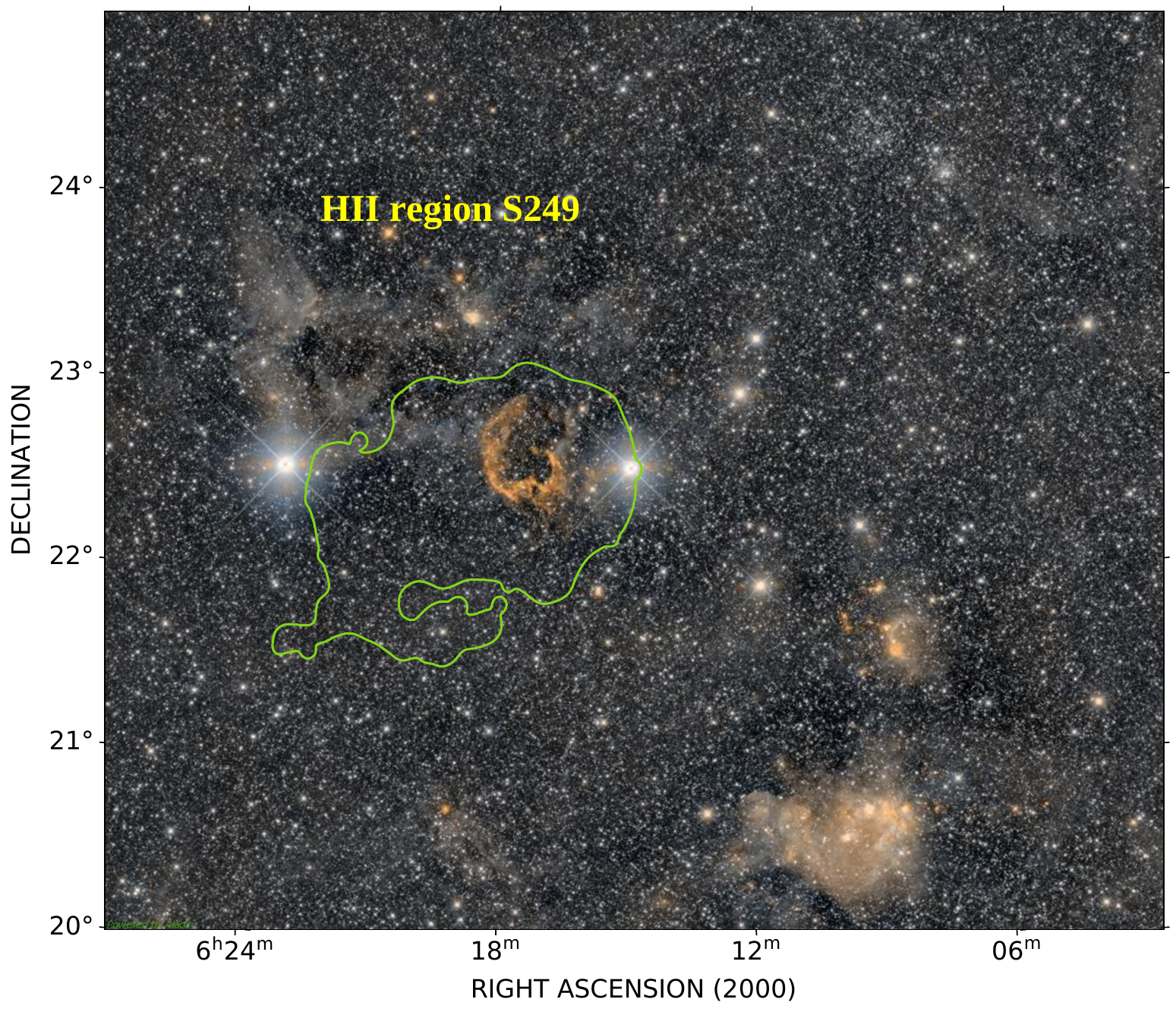}
    \caption{Overplot of the X-ray contours obtained with eROSITA (light green) with the un\textit{WISE} color archival data (W2, 4.6 $\mu$m; W1, 3.4$\mu $m).
    \label{fig:WISE_overplot}}
\end{figure}

\footnotetext[\thefootnote]{Identified with the source Gaia EDR3 3376739988615000320 and AllWISE J061853.77+214551.5}
 In order to better visualize the amount of material in the region, we over plotted the X-ray contours of our observation to the \textit{WISE} archival data. The result is shown in Figure \ref{fig:WISE_overplot}: the emission of G189.6+03.3 partially overlaps with the nearby S249 HII region, which is bright in this image. We recall how \cite{Fesen1984} showed this region is interacting with IC443.

\begin{figure}
    \centering
    \includegraphics[scale=0.5]{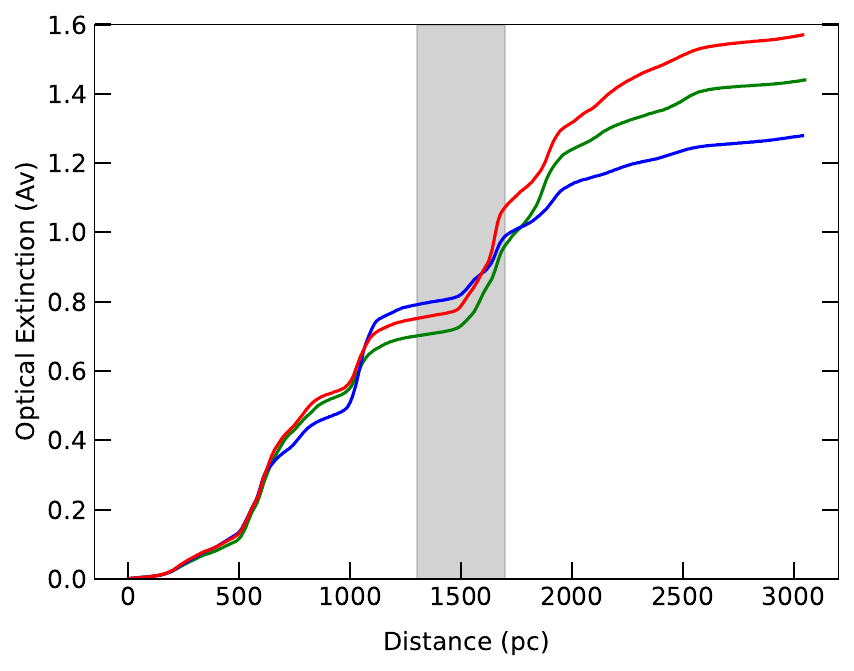}
    \caption{Optical extinction in three different directions. \textit{Red}: IC443. \textit{Green}: G189.6+03.3.  \textit{Blue}: HII region S249. The dataset have been obtained from the Gaia/2MASS extinction map available at \url{https://astro.acri-st.fr/gaia_dev/}. We also report the distance of $1.5\pm 0.2$ pc adopted in the paper \citep{Fesen1984,Welsh2003}.
    \label{fig:Gaia_extinction}}
\end{figure}

Observing Figure \ref{fig:WISE_overplot}, we expected different absorption values across G189.6+03.3 and IC443. Therefore, we looked at the optical extinction data provided by \cite{Lallement2019} and available online\footnote{\url{https://astro.acri-st.fr/gaia_dev/}}. This database uses the parallax derived distance of \textit{Gaia} and the optical extinction measured with the same instrument to estimate the distance of the dust. In Figure \ref{fig:Gaia_extinction} are shown the extinction data from three different spots, one from the central region of G189.6+03.3, one in the direction of IC443 and one in the direction of the HII region S249 (indicated in Figure \ref{fig:WISE_overplot}). The optical extinction curves are quite similar to each other, indicating that the three regions are absorbed by the same amount of dust and hence are likely located at similar distance from us. It is however unclear whether the arc like structure visible in optical in the North is compressed material by the shockwave originating from G189.6+03.3 (as proposed by \cite{Asaoka1994}) or if it is still part of IC443 \citep{Fesen1984}. Even though optical extinction can be related to X-ray column density value \citep[see for example][]{Predehl1995}, we notice how deviations from this formula can easily be justified by dust being destroyed by the blast wave during the supernova explosion \citep{Micelotta2016,Zhu2019}. In addition, uncertainties in the \textit{Gaia} optical extinction measurements increase above 2kpc. Therefore, we present the profiles in Figure \ref{fig:Gaia_extinction} only for a qualitative comparison.

\section{Spectral Analysis\label{sec:spectral analysis}}
The analysis of the X-ray spectra has been carried out with PyXSPEC, the Python interface of XSPEC \citep{Xspec}, and the errors are expressed in 1$\sigma$ confidence level. We employed the Cash statistics \citep{Cash1979}, using the version implemented in XSPEC. The choice of using the ratio CSTAT/dof (dof=degree of freedom) to estimate the goodness of the fit is motivated by the fact that for sufficient number of counts, the C statistic approaches the $\chi^{2}$ statistic \citep{Kaastra2017}. In order to estimate the errors in a robust way, we run an MCMC (Monte Carlo Markov Chain) based code using the Python library \texttt{emcee} \citep{emcee2013}, running it for 40000 steps. We considered only the last 2000 steps of the run with the purpose to get the largest possible number of chains converged. We initialized our walkers with a Gaussian distribution centered on the best fit parameters, employing logarithmically uniform priors on the model components which were left free to vary during the run. One of the main advantage of this approach over the traditional fitting technique is the capability to probe more in depth the space of parameters \citep[see for example][for a description of the advantange of using an MCMC based approach in X-ray astronomy]{vanDyk2001}. We extracted the spectra from the regions shown in Figure \ref{fig:G189_IC443_RGB}. Before proceeding with the spectral analysis, we removed the point sources in both background and source regions.

\begin{figure*}
    \centering
    \includegraphics[width=6.5cm]{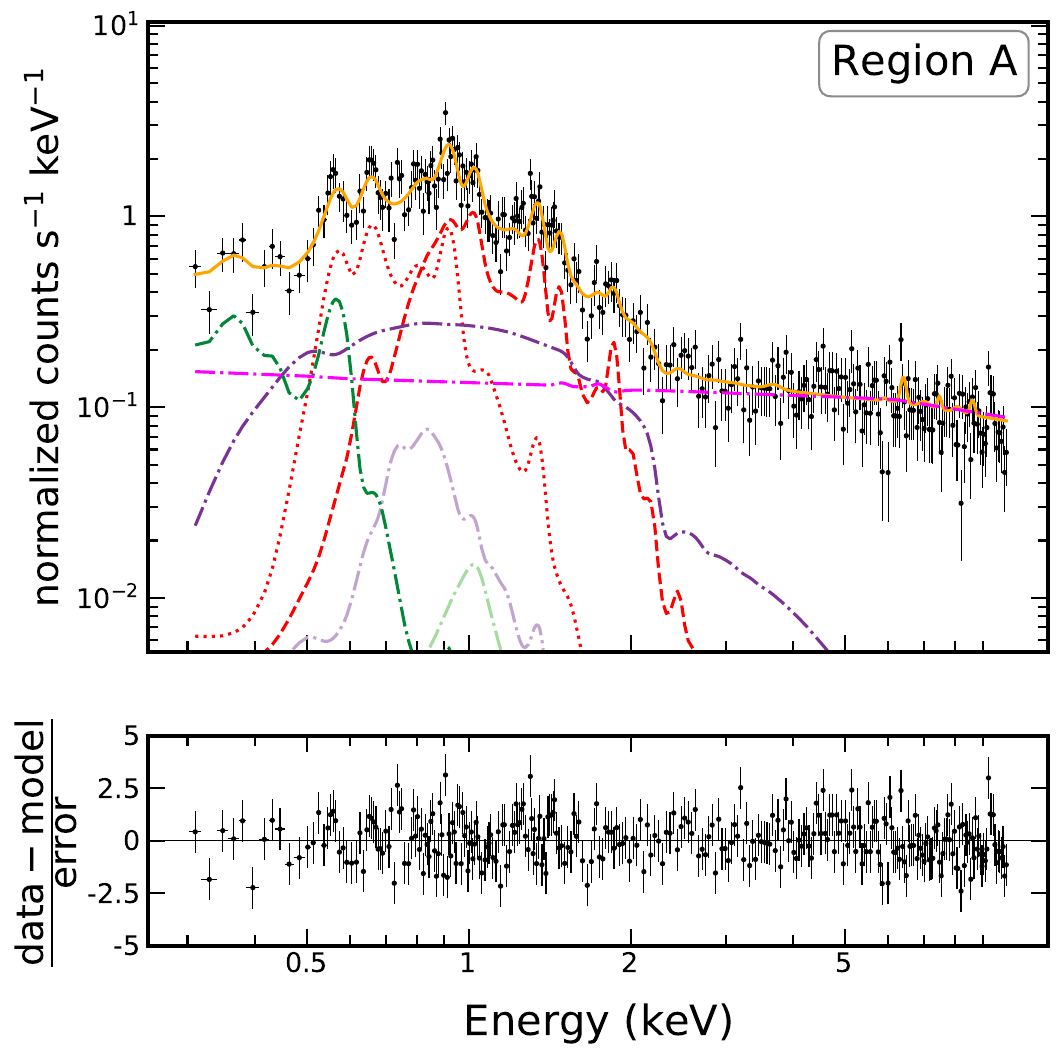}
    \includegraphics[width=6.5cm]{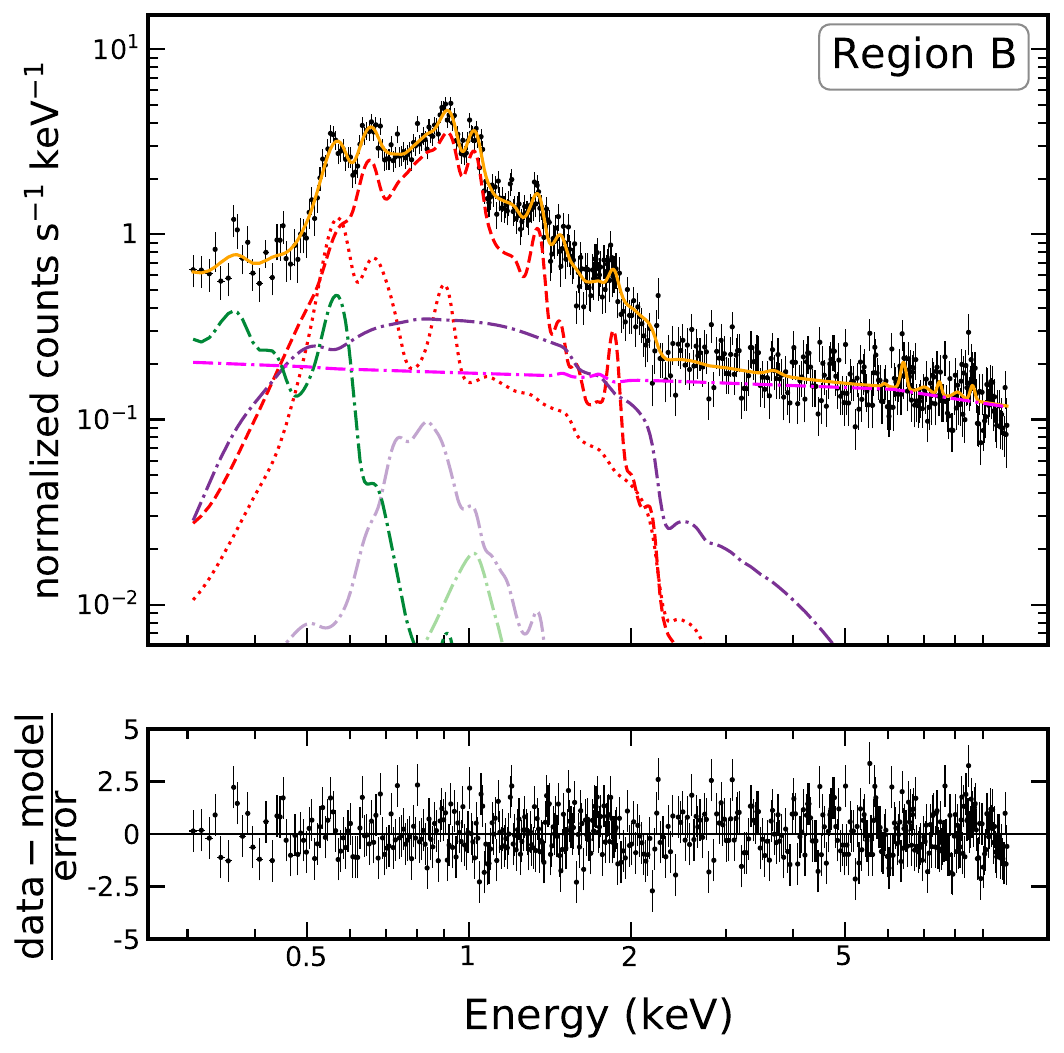}
    \includegraphics[width=6.5cm]{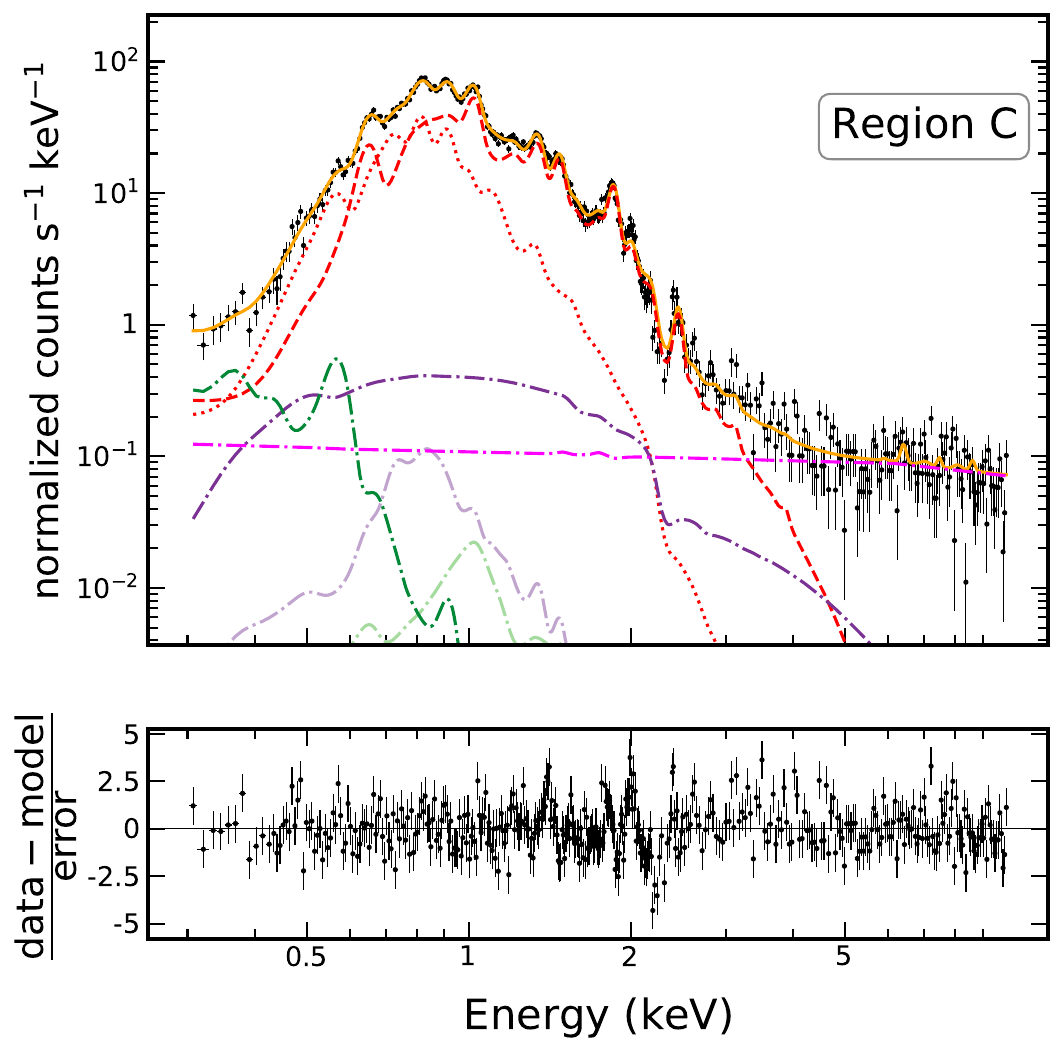}
    \includegraphics[width=6.5cm]{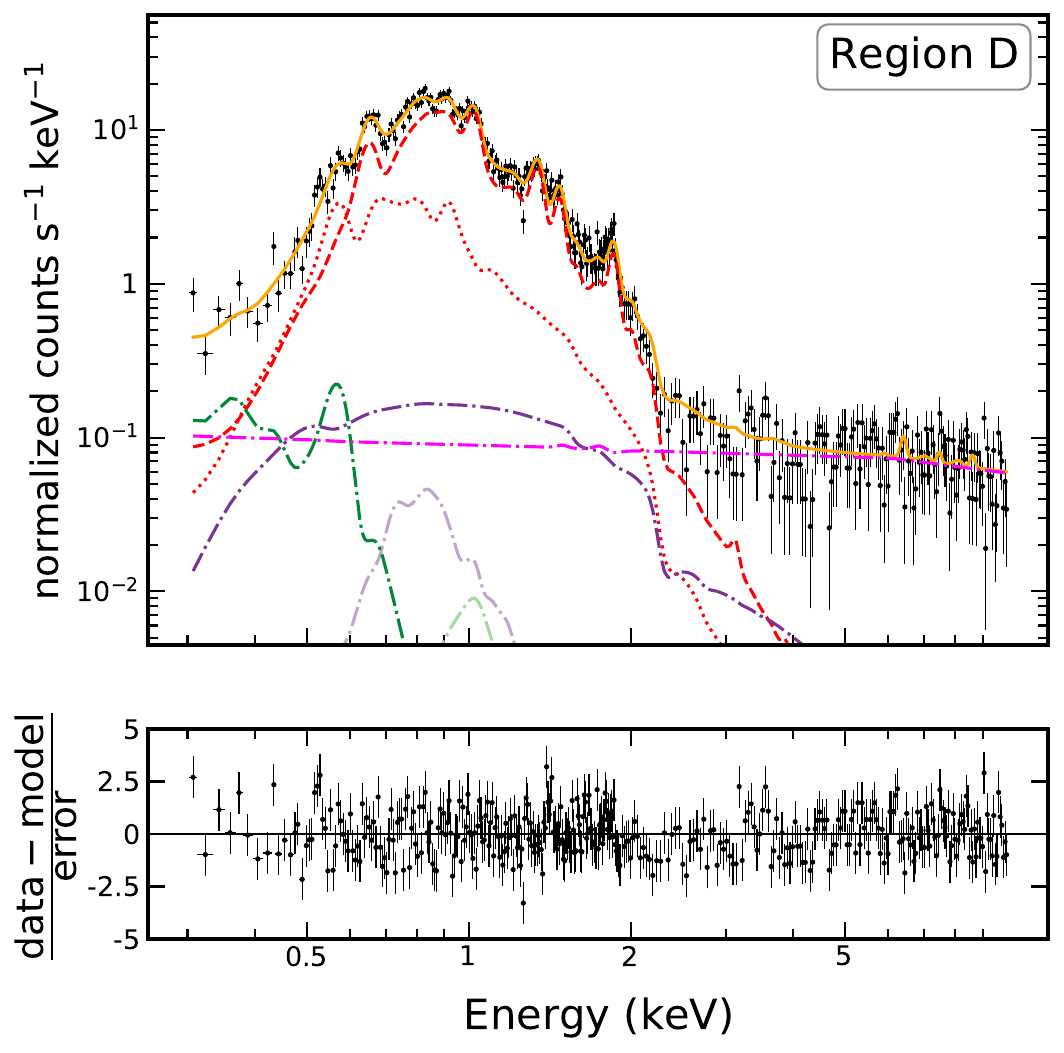}
    \caption{Spectra from region A, B, C and D. The solid yellow line indicates the total model, the red dotted line represents the first additive component (VPSHOCK), the red dashed line the second additive component (VPSHOCK). The dashdot lines show the different background components: the horizontal magenta line is the most significant component of the particle background, the others represent the sky background made by the Local Hot Bubble (dark green), Cosmic X-ray Background (violet), Cosmic Halo 1 (light violet) and Cosmic Halo 2 (light green). The spectra have been rebinned for displaying purpose and the parameters of the model are the median values of the last 2000 steps of the \texttt{emcee} run.
    \label{fig:G189_IC443_spectra}}
\end{figure*}

 We started modeling the background spectrum following the approach of \cite{Okon2021} and references therein. The background extraction region is indicated in Figure \ref{fig:G189_IC443_RGB}. The model consists of one powerlaw component representing the Cosmic X-ray Background with a fixed slope of 1.4 and three collisional equilibrium thermal model \citep[APEC,][]{Smith2001} components, each one with fixed temperature and solar abundances, but free normalization. The Local Hot Bubble is modeled with temperature of 0.105 keV, while the Galactic Halo is described by the other two APEC models respectively with $kT=0.658$ keV and $kT=1.22$ keV. We absorbed the Galactic Halo and Cosmic X-ray Background components with the TBabs model \citep{Wilms2000}. Given this modeling, we determined the best fitting background model, first running a fit on the background region. We then proceed fitting the background best fitting model simultaneously with the source model to the spectrum of each one of the regions shown in orange in Figure \ref{fig:G189_IC443_RGB}. Since from Figure \ref{fig:G189_IC443_RGB} the background is not significantly variable across the regions, we fixed the shape of the background model, fitting only a global normalization parameter simultaneously with the source. As an additional test, we checked our results employing another background region extracted few degrees South of G189.6+03.3, obtaining the same results. The spectra are shown in Figure \ref{fig:G189_IC443_spectra}.

 Following the labeling in Figure \ref{fig:G189_IC443_RGB}, we started fitting region 'A' with a constant temperature plane parallel shock thermal model \citep[VPSHOCK,][]{Borkowski2001}, absorbed with the model TBabs \citep{Wilms2000}. This region should contain only the contribution from G189.6+03.3. The goodness of the fit was not acceptable, so that we decided to add another thermal shock component (VPSHOCK) applying also a velocity shift (VASHIFT) to the initial model. We were motivated by the fact that from Figure \ref{fig:G189_IC443_RGB} the inner region 'A' looks surrounded by a brighter emission, which we analyzed separately as region 'B'. With the two components model, we assume one component represents the inner emission of the remnant and the other one the emission from an expanding shell. In the expanding component (the one multiplied by the velocity shift) we left as free abundances only O, Mg, Ne and Fe, assuming that these elements are those mainly enriched, given also that are associated to the brightest lines. In the same component we have frozen to 0 the abundance values of C, N, Si and S to avoid excessive degeneracy with the parameters of the other component. In the rest component we instead set free to vary C, N, O, Mg, Ne, Fe, Si and S: this model definitely improves the fit statistic (see Table \ref{tab:BF_2VPSHOCK} for the goodness of fits). The abundances not mentioned above are left frozen at solar value.

 As mentioned in Section \ref{sec:intro}, starting from \cite{Yamaguchi2009} recombination was clearly detected in the spectra of IC443 by several studies, while it was found as a relevant process also in G189.6+03.3 \citep{Yamauchi2020}. Therefore, we also tested a recombining plasma model \citep[VRNEI,][]{Foster2017} as a second additive component. The abundances were set as with the double VPSHOCK model and the initial cooling temperature T$_{0}$ is set frozen to 5 keV. We decided to make the same assumption done by \cite{Okon2021}: in this way, the O-S elements are fully ionized in the initial condition. Looking at the spectra, no extra residuals appear when a two component model is applied and the difference of the statistics (Table \ref{tab:BF_2VPSHOCK},\ref{tab:BF_VRNEI}) between two VPSHOCK models and a single VPSHOCK plus a recombination component (VRNEI) is minimal. The ionization timescale parameter in both models is close to 10$^{12}$ cm$^{-3}$, indicating the gas is close to the ionization equilibrium.  However, the column density is significantly higher in the double VPSHOCK model: we tried to investigate this feature, freezing some parameters in the velocity shifted VPSHOCK component to the same values of the same parameters in the VPSHOCK+VRNEI model. We found out that the double VPSHOCK retrieves a column density ($0.50_{-0.09}^{+0.10}\cdot 10^{22}$ cm$^{-2}$) compatible with $0.41_{-0.11}^{+0.10}\cdot 10^{22}$ cm$^{-2}$ obtained with VPSHOCK+VRNEI if the ionization timescale and velocity shift are set equal to the values found in the VPSHOCK of the recombination model. We tried to add also a black body radiation model to model the faint emission from the inner part (10 km as diameter, distance of 1500 pc, 0.1 keV temperature) but the fit did not improve significantly.

In region 'B' we analyze the bright external part of G189.6+03.3 emission, possibly associated to a shell. We initially tested a single VPSHOCK model, but the fit retrieved strong residuals. The spectra are again better described by a two component shocked plasma. Either we use a VPSHOCK or VRNEI as a second component, the column density is always $\sim 4.0\cdot10^{21}$ cm$^{-2}$ with similar statistic values. If we use VPSHOCK as a second component on top of the first VPSHOCK, we find a first plasma component with temperature $kT=2.3^{+2.3}_{-0.1}$ keV, with the second component showing $kT=0.74^{+0.07}_{-0.08}$ keV. 
From Table \ref{tab:BF_2VPSHOCK}, the double VPSHOCK model provides a CSTAT/dof ratio close to 1, making it the best fit comparing to all tests we carried out. Therefore, we find an additional evidence for our initial hypothesis: the hot 2.3 keV expanding component can be associated to an expanding shell, while the inner emission is cooler with a temperature 0.7 keV. In Section \ref{sec:Spatial Analysis} we observed how the north part of G189.6+03.3 is coincident with a dust structure visible in \textit{WISE} data: it seems therefore reasonable to argue the enhancement in the temperature of the plasma might be given by the compression of the shock against a denser medium. This would eventually imply that G189.6+03.3 and the HII region S249 are at the same distance.  We also observe that the ionization timescale in the 0.7 keV component for the double VPSHOCK model is $4\cdot 10^{11}$ cm s$^{-3}$ while in VPSHOCK plus VRNEI is $3\cdot 10^{12}$ cm s$^{-3}$: since the statistics improves for the double VPSHOCK, we suggest the plasma is hot and recently shocked in this region.  The higher value of $\tau$ in region 'B' also in the 0.7 keV component comparing to the other regions (Table \ref{tab:BF_2VPSHOCK},\ref{tab:BF_VRNEI}) is probably a consequence of the recent interaction of the gas with the nearby HII region. 

\begin{table*}[]
\centering
\caption{Results from the spectral fits of the different regions using  a double VPSHOCK model.\label{tab:BF_2VPSHOCK}}.
\begin{tabular}{ccccc}
Model&tbabs*(vashift*vpshock+vpshock)&&&\\[1ex]%
\hline%
Region&A&B&C&D\\[1ex]%
\hline%
factor&$0.111_{-0.002}^{+0.003}$&$0.155_{-0.003}^{+0.003}$&$0.095_{-0.003}^{+0.004}$&$0.079_{-0.002}^{+0.003}$\\[1ex]%
N$\_$H (10$^{22}$ cm$^{-2}$)&$0.82_{-0.15}^{+0.14}$&$0.37_{-0.06}^{+0.06}$&$0.613_{-0.020}^{+0.014}$&$0.44_{-0.03}^{+0.03}$\\[1ex]%
Velocity (km/s)&$0_{-0}^{+500}$&$4400_{-2100}^{+1900}$&$2600_{-500}^{+500}$&$0_{-0}^{+2800}$\\[1ex]%
kT (keV)&$0.174_{-0.012}^{+0.014}$&$2.3_{-1.0}^{+2.3}$&$0.30_{-0.03}^{+0.03}$&$0.43_{-0.06}^{+0.08}$\\[1ex]%
O/O$_{\odot}$&$3_{-2}^{+2}$&$0.6_{-0.3}^{+0.4}$&$0.049_{-0.007}^{+0.012}$&$0.08_{-0.02}^{+0.03}$\\[1ex]%
Ne/Ne$_{\odot}$&$3.3_{-1.5}^{+1.5}$&$0.17_{-0.14}^{+0.26}$&$0.18_{-0.03}^{+0.03}$&$0.16_{-0.10}^{+0.11}$\\[1ex]%
Mg/Mg$_{\odot}$&$2_{-2}^{+5}$&$0.04_{-0.02}^{+0.06}$&$0.12_{-0.09}^{+0.14}$&$0.03_{-0.01}^{+0.07}$\\[1ex]%
Fe/Fe$_{\odot}$&$5_{-3}^{+4}$&$1_{-1}^{+4}$&$0.9_{-0.3}^{+0.3}$&$4_{-2}^{+3}$\\[1ex]%
$\tau_{u}$ (10$^{10}$ cm$^{-3}$ s)&$120_{-80}^{+270}$&$0.25_{-0.09}^{+0.24}$&$2.6_{-0.6}^{+1.4}$&$0.7_{-0.2}^{+0.3}$\\[1ex]%
Normalization &$0.12_{-0.05}^{+0.08}$&$0.014_{-0.004}^{+0.007}$&$8.0_{-2.1}^{+1.4}$&$0.6_{-0.2}^{+0.2}$\\[1ex]%
kT (keV)&$0.85_{-0.17}^{+0.17}$&$0.74_{-0.08}^{+0.07}$&$0.775_{-0.016}^{+0.014}$&$0.74_{-0.03}^{+0.02}$\\[1ex]%
C/C$_{\odot}$&$1_{-1}^{+6}$&$0.03_{-0.02}^{+0.04}$&$0.06_{-0.04}^{+0.24}$&$0.2_{-0.2}^{+3.0}$\\[1ex]%
N/N$_{\odot}$&$0.017_{-0.006}^{+0.018}$&$0.04_{-0.03}^{+0.12}$&$1_{-1}^{+4}$&$0.05_{-0.03}^{+0.22}$\\[1ex]%
O/O$_{\odot}$&$3_{-3}^{+5}$&$2.1_{-1.3}^{+1.9}$&$7_{-2}^{+2}$&$5_{-2}^{+3}$\\[1ex]%
Ne/Ne$_{\odot}$&$6_{-3}^{+3}$&$4_{-2}^{+3}$&$3.7_{-1.1}^{+0.8}$&$2.8_{-0.8}^{+1.8}$\\[1ex]%
Mg/Mg$_{\odot}$&$3.7_{-1.6}^{+2.3}$&$2.7_{-1.1}^{+1.8}$&$1.8_{-0.4}^{+0.3}$&$1.9_{-0.5}^{+1.0}$\\[1ex]%
Si/Si$_{\odot}$&$1.5_{-1.0}^{+1.3}$&$2.4_{-1.0}^{+1.9}$&$1.4_{-0.4}^{+0.3}$&$1.0_{-0.3}^{+0.4}$\\[1ex]%
S/S$_{\odot}$&$1_{-1}^{+4}$&$0.04_{-0.02}^{+0.06}$&$1.5_{-0.4}^{+0.4}$&$0.04_{-0.03}^{+0.09}$\\[1ex]%
Fe/Fe$_{\odot}$&$0.4_{-0.4}^{+1.0}$&$1.2_{-0.6}^{+0.9}$&$0.41_{-0.09}^{+0.09}$&$0.48_{-0.13}^{+0.22}$\\[1ex]%
$\tau_{u}$ (10$^{10}$ cm$^{-3}$ s)&$110_{-70}^{+210}$&$40_{-11}^{+27}$&$600_{-200}^{+300}$&$400_{-200}^{+300}$\\[1ex]%
Normalization & $0.011_{-0.004}^{+0.006}$&$0.008_{-0.003}^{+0.006}$&$1.0_{-0.2}^{+0.3}$&$0.22_{-0.07}^{+0.07}$\\[1ex]%
\hline%
Background model& & & &\\[1ex]%
\hline%
factor&$0.47_{-0.04}^{+0.05}$&$0.84_{-0.10}^{+0.09}$&$0.28_{-0.08}^{+0.09}$&$0.22_{-0.06}^{+0.07}$\\[1ex]%
\hline
Statistic&927/865&868/865&1110/865&1077/865\\[1ex]%
\end{tabular}
\begin{flushleft}
\small
 Normalization is expressed as $10^{-14}\dfrac{\int n_{e}n_{H}dV}{4\pi D^{2}}$ where n$_{e}$ is the electron density of the plasma (cm$^{-3}$), n$_{H}$ is the hydrogen density (cm$^{-3}$) and D (cm) is the distance of the source
\end{flushleft}
\end{table*}
\begin{table*}[]
\centering
\caption{Results from the spectral fits of the different regions using a VPSHOCK plus VRNEI model. \label{tab:BF_VRNEI}}
\begin{tabular}{ccccc}
\hline%
Model&tbabs*(vashift*vpshock+vrnei)&&&\\[1ex]%
\hline%
Region&A&B&C&D\\[1ex]%
\hline%
factor&$0.111_{-0.003}^{+0.003}$&$0.156_{-0.002}^{+0.003}$&$0.091_{-0.003}^{+0.004}$&$0.078_{-0.003}^{+0.004}$\\[1ex]%
N$\_$H (10$^{22}$ cm$^{-2}$)&$0.41_{-0.11}^{+0.10}$&$0.38_{-0.07}^{+0.05}$&$0.71_{-0.03}^{+0.03}$&$0.32_{-0.02}^{+0.02}$\\[1ex]%
Velocity (km/s)&$0_{-0}^{+160}$&$800_{-800}^{+1600}$&$2300_{-800}^{+600}$&$0_{-0}^{+500}$\\[1ex]%
kT (keV)&$1.6_{-0.7}^{+1.7}$&$0.8_{-0.2}^{+0.5}$&$0.25_{-0.03}^{+0.03}$&$0.77_{-0.02}^{+0.03}$\\[1ex]%
O/O$_{\odot}$&$0.7_{-0.4}^{+1.1}$&$1.2_{-0.4}^{+0.5}$&$0.24_{-0.08}^{+0.11}$&$2.9_{-1.1}^{+1.6}$\\[1ex]%
Ne/Ne$_{\odot}$&$2.0_{-0.8}^{+1.6}$&$1.5_{-0.5}^{+0.9}$&$0.8_{-0.2}^{+0.3}$&$1.7_{-0.5}^{+1.1}$\\[1ex]%
Mg/Mg$_{\odot}$&$5_{-4}^{+3}$&$1.6_{-0.8}^{+1.2}$&$0.1_{-0.1}^{+0.4}$&$2.4_{-0.8}^{+0.9}$\\[1ex]%
Fe/Fe$_{\odot}$&$4_{-4}^{+3}$&$0.5_{-0.4}^{+0.5}$&$7_{-3}^{+2}$&$1.6_{-0.6}^{+0.5}$\\[1ex]%
$\tau_{u}$ (10$^{10}$ cm$^{-3}$ s)&$0.41_{-0.11}^{+0.16}$&$2.6_{-1.0}^{+2.3}$&$2.3_{-0.7}^{+0.9}$&$49_{-15}^{+20}$\\[1ex]%
Normalization &$0.004_{-0.002}^{+0.003}$&$0.008_{-0.003}^{+0.002}$&$4.3_{-1.6}^{+1.8}$&$0.07_{-0.02}^{+0.04}$\\[1ex]%
kT (keV)&$0.76_{-0.16}^{+0.14}$&$0.72_{-0.10}^{+0.10}$&$0.416_{-0.019}^{+0.019}$&$0.50_{-0.05}^{+0.05}$\\[1ex]%
C/C$_{\odot}$&$0.03_{-0.01}^{+0.04}$&$0.07_{-0.05}^{+0.24}$&$0.3_{-0.3}^{+2.4}$&$1_{-1}^{+6}$\\[1ex]%
N/N$_{\odot}$&$0.019_{-0.007}^{+0.020}$&$0.03_{-0.02}^{+0.07}$&$5_{-2}^{+2}$&$0.03_{-0.02}^{+0.06}$\\[1ex]%
O/O$_{\odot}$&$5_{-2}^{+3}$&$0.05_{-0.03}^{+0.16}$&$1.0_{-0.2}^{+0.3}$&$0.2_{-0.2}^{+0.7}$\\[1ex]%
Ne/Ne$_{\odot}$&$3.6_{-1.2}^{+2.5}$&$2.8_{-1.2}^{+2.8}$&$0.87_{-0.15}^{+0.18}$&$5.4_{-1.3}^{+1.8}$\\[1ex]%
Mg/Mg$_{\odot}$&$2.3_{-0.8}^{+1.7}$&$0.6_{-0.4}^{+0.6}$&$0.88_{-0.12}^{+0.16}$&$0.06_{-0.04}^{+0.19}$\\[1ex]%
Si/Si$_{\odot}$&$0.9_{-0.7}^{+1.2}$&$1.0_{-0.4}^{+0.9}$&$1.28_{-0.17}^{+0.21}$&$0.03_{-0.02}^{+0.05}$\\[1ex]%
S/S$_{\odot}$&$1_{-1}^{+5}$&$0.2_{-0.1}^{+0.8}$&$1.1_{-0.2}^{+0.3}$&$0.03_{-0.01}^{+0.05}$\\[1ex]%
Fe/Fe$_{\odot}$&$0.2_{-0.1}^{+0.3}$&$0.14_{-0.12}^{+0.19}$&$0.18_{-0.03}^{+0.05}$&$0.024_{-0.010}^{+0.023}$\\[1ex]%
$\tau$ (10$^{10}$ cm$^{-3}$ s)&$200_{-100}^{+300}$&$300_{-200}^{+500}$&$56_{-2}^{+2}$&$0.06_{-0.05}^{+0.35}$\\[1ex]%
Normalization &$0.014_{-0.005}^{+0.005}$&$0.029_{-0.011}^{+0.014}$&$5.4_{-1.1}^{+0.9}$&$0.44_{-0.10}^{+0.10}$\\[1ex]%
\hline%
Background model& & & &\\[1ex]%
\hline%
factor&$0.43_{-0.06}^{+0.06}$&$0.88_{-0.10}^{+0.11}$&$0.38_{-0.07}^{+0.08}$&$0.20_{-0.08}^{+0.07}$\\[1ex]%
\hline
Statistic&931/865&880/865&972/865&1184/865\\[1ex]%
\end{tabular}
\begin{flushleft}
\small
 Normalization is expressed as $10^{-14}\dfrac{\int n_{e}n_{H}dV}{4\pi D^{2}}$ where n$_{e}$ is the electron density of the plasma (cm$^{-3}$), n$_{H}$ is the hydrogen density (cm$^{-3}$) and D (cm) is the distance of the source
\end{flushleft}
\end{table*}

We then moved to analyze the emission of regions 'C' and 'D' to understand whether the plasma emission of G189.6+03.3 overlaps with IC443, as initially proposed by \cite{Asaoka1994}. Indeed, these two regions are those covering IC443. For region 'C' and 'D'  we find as in region 'B' a double component model fits best the data. As visible in Table \ref{tab:BF_2VPSHOCK}, if VPSHOCK is employed as second component, we detect a temperature close to $kT\sim$ 0.7 keV in both regions. Conversely, in region 'C' employing VRNEI as second component the fits retrieve a considerable improved statistics, but do not show hints of the 0.7 keV component (Table \ref{tab:BF_VRNEI}). Moreover, also the column density is not consistent with $4.0\cdot10^{21}$ cm$^{-2}$, a value found in several other regions. Region 'C' also displays a considerably high expansion velocity which is not detected in region 'D', which was supposed to belong to the same structure in the current literature. However, a justification for this difference is that the shock has been slowed down by the interaction with a molecular cloud in region D \citep{Cornett1977,Ustamujic2021}. Observing Figure \ref{fig:G189_IC443_RGB}, the two regions have similar surface brightness, which is quite high especially in region 'C': therefore is possible that the dim 0.7 keV component is not resolved in the spectrum of this region. We tested this scenario, adding an APEC component to the models employed before. The idea behind this choice is that the material is contained by the molecular cloud, possibly reheated by the reverse shock generated by the shock wave impacting on the molecular cloud itself. The results are shown in Table \ref{tab:BF_APEC_ADD}. 

With the addition of the APEC model to the VRNEI+VPSHOCK model, we obtain an improved statistic ($\Delta$ CSTAT=23) and the 0.7 keV component appears again. Considering the double VPSHOCK description  provides a considerably worse statistic, our conclusion is that the relaxed material (APEC) here detected belongs to G189.6+03.3 while the VSHOCK+VRNEI component is the shocked emission coming from IC443. This confirms the previous results that found recombination and overionized material from this region \citep{Yamaguchi2009,Greco2018}.

\begin{table*}[]
\centering
\caption{Results from the spectral fits of region 'C' using an additional APEC component on top of the model employed before. \label{tab:BF_APEC_ADD}}
\begin{tabular}{ccc}
\hline%
Region&C&\\[1ex]%
\hline%
Model&tbabs*(vashift*vpshock+vrnei+apec)&tbabs*(vashift*vpshock+vpshock+apec)\\[1ex]%
\hline%
factor&$0.091_{-0.003}^{+0.004}$&$0.093_{-0.004}^{+0.004}$\\[1ex]%
N$\_$H (10$^{22}$ cm$^{-2}$)&$0.75_{-0.02}^{+0.03}$&$0.55_{-0.03}^{+0.03}$\\[1ex]%
Velocity (km/s)&$1600_{-800}^{+800}$&$3000_{-600}^{+500}$\\[1ex]%
kT (keV)&$0.25_{-0.02}^{+0.02}$&$0.217_{-0.018}^{+0.019}$\\[1ex]%
O/O$_{\odot}$&$0.4_{-0.1}^{+0.3}$&$0.4_{-0.2}^{+0.4}$\\[1ex]%
Ne/Ne$_{\odot}$&$0.6_{-0.2}^{+0.3}$&$1.6_{-0.5}^{+0.7}$\\[1ex]%
Mg/Mg$_{\odot}$&$2.8_{-1.2}^{+1.1}$&$0.05_{-0.03}^{+0.12}$\\[1ex]%
Fe/Fe$_{\odot}$&$7_{-2}^{+2}$&$8_{-2}^{+2}$\\[1ex]%
$\tau$ (10$^{10}$ cm$^{-3}$ s)&$3.9_{-1.3}^{+1.8}$&$13_{-5}^{+13}$\\[1ex]%
Normalization&$3.1_{-1.1}^{+1.3}$&$1.1_{-0.5}^{+0.6}$\\[1ex]%
kT (keV)&$0.288_{-0.013}^{+0.014}$&$0.76_{-0.02}^{+0.02}$\\[1ex]%
C/C$_{\odot}$&$7_{-5}^{+2}$&$1_{-1}^{+5}$\\[1ex]%
N/N$_{\odot}$&$6_{-2}^{+3}$&$7_{-2}^{+2}$\\[1ex]%
O/O$_{\odot}$&$1.2_{-0.3}^{+0.4}$&$3.0_{-0.5}^{+0.7}$\\[1ex]%
Ne/Ne$_{\odot}$&$2.0_{-0.4}^{+0.6}$&$1.9_{-0.3}^{+0.4}$\\[1ex]%
Mg/Mg$_{\odot}$&$0.86_{-0.16}^{+0.21}$&$1.12_{-0.14}^{+0.19}$\\[1ex]%
Si/Si$_{\odot}$&$1.5_{-0.3}^{+0.3}$&$0.88_{-0.10}^{+0.15}$\\[1ex]%
S/S$_{\odot}$&$0.9_{-0.2}^{+0.3}$&$0.9_{-0.2}^{+0.3}$\\[1ex]%
Fe/Fe$_{\odot}$&$0.02_{-0.01}^{+0.03}$&$0.32_{-0.04}^{+0.05}$\\[1ex]%
$\tau$ (10$^{10}$ cm$^{-3}$ s)&$39_{-3}^{+3}$&$270_{-60}^{+90}$\\[1ex]%
Normalization&$4.6_{-1.0}^{+0.8}$&$1.42_{-0.19}^{+0.15}$\\[1ex]%
kT (keV)&$0.66_{-0.04}^{+0.03}$&$1.5_{-0.6}^{+0.4}$\\[1ex]%
Abundance (Z$_{\odot}$)&$1.1_{-0.3}^{+0.4}$&$1.3_{-0.9}^{+1.2}$\\[1ex]%
Normalization&$0.50_{-0.13}^{+0.22}$&$0.09_{-0.04}^{+0.28}$\\[1ex]%
\hline%
Background model & &\\[1ex]%
\hline%
factor&$0.37_{-0.07}^{+0.09}$&$0.32_{-0.09}^{+0.10}$\\[1ex]%
\hline
Statistic&949.29/862&1107.45/862\\[1ex]%
\end{tabular}
\begin{flushleft}
\small
 Normalization is expressed as $10^{-14}\dfrac{\int n_{e}n_{H}dV}{4\pi D^{2}}$ where n$_{e}$ is the electron density of the plasma (cm$^{-3}$), n$_{H}$ is the hydrogen density (cm$^{-3}$) and D (cm) is the distance of the source
\end{flushleft}
\end{table*}

Looking at region 'D', we find the 0.7 keV component with ionization timescale factor of the plasma $\tau=400_{-200}^{+300}\cdot 10^{10}$ cm s$^{-3}$, again giving strongly evidence for a plasma close to the ionization equilibrium with physical properties very similar to those found in the other regions. The very low speed detected is an additional proof that this relaxed gas was probably slowed down in the past due to the interaction with a nearby molecular cloud.

We recall how $\tau=10^{13}$ s cm$^{-3}$ is assumed as the upper limit of the ionization timescale parameter, implying collisional equilibrium in the VPSHOCK model \citep{Xspec,Borkowski1994,Borkowski2001}. 

In conclusion, in all the regions we find a constant temperature plasma component $kT=0.7$ keV, suggesting the emission of G189.06+3.3 covers all the regions analyzed, including those whose emission is associated to IC443 (region C and region D).


\subsection{An unresolved source at the center of G189.6+03.3? \label{sec:diffuse_emission_center_G189}}

\begin{figure}
    \centering
    \includegraphics[scale=0.5]{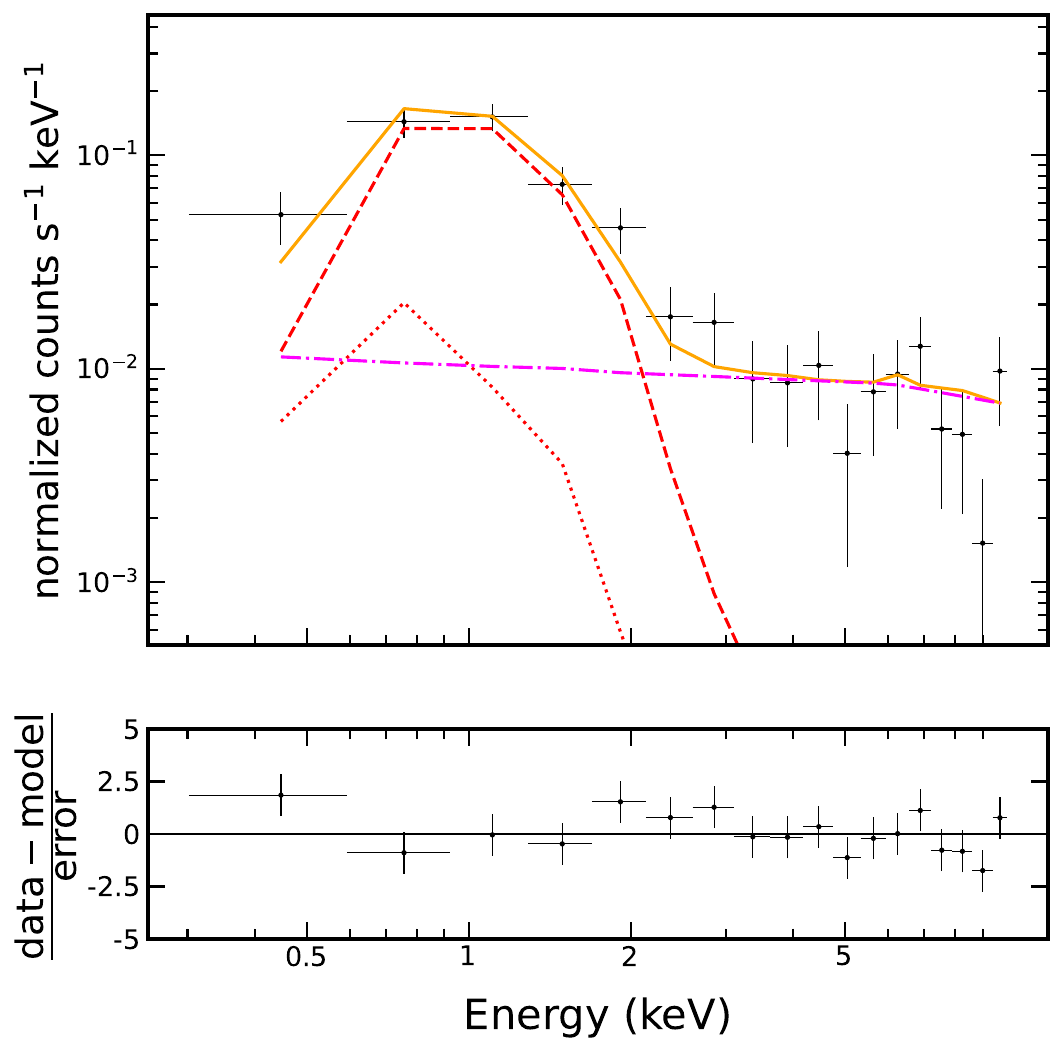}
    \caption{Spectrum of the diffuse emission close to the center of G189.6+03.3 (light blue circle in Figure \ref{fig:G189_IC443_zoom}) modeled with the VPSHOCK+VRNEI model. The data have been rebinned for displaying purpose. The two red lines indicate the two model components, while the magenta represents the instrumental background which we modeled. 
    \label{fig:PWN_candidate_spec}}
\end{figure}

Employing the best fit obtained above and despite the very few counts available, we modeled the emission of the diffuse emission close to the center of G189.6+03.3 described in Section \ref{sec:Spatial Analysis}. The spectra are background subtracted, except for the instrumental component that we continued to model separately to describe the remaining high energy tail. We first tested an unabsorbed POWERLAW with photon index fixed to 2 and free normalization to derive a flux in 0.2-10 keV band from the light blue circle indicated in Figure \ref{fig:G189_IC443_zoom}. We left free to vary also the column density, obtaining 0.33$\cdot10^{22}$ cm$^{-2}$. The unabsorbed background subtracted flux is 9.38$_{-1.18}^{+1.62} \cdot 10^{-13}$ erg s$^{-1}$ cm$^{-2}$. For the same region, we also tested the best fit VPSHOCK+VRNEI model of region A (as discussed in Section \ref{sec:spectral analysis} this has been proven to be more reliable than double VPSHOCK), obtaining an unabsorbed background subtracted flux of 5.19$_{-0.57}^{+1.29} \cdot 10^{-13}$ erg s$^{-1}$ cm$^{-2}$. The spectrum is shown in Figure \ref{fig:PWN_candidate_spec}. Despite we find a similar value of the CSTAT/DOF ratio (0.59) for both models, we find a considerable higher flux with the powerlaw model at 90\% confidence level. Given its extended nature and the position almost at the center of G189.6+03.3, the object could be a pulsar wind nebula (PWN). Therefore, deeper observations with \textit{Chandra} or \textit{XMM-Newton} are needed to assess the nature of the object, also considering that no radio detection is associated to this position.

\subsection{Two star clusters in the neighborhood \label{sec:open_clusters}}
Thanks to the almost unlimited field of view of eROSITA provided by the all-sky scan survey mode, we were able to image two open clusters in the same sky region of IC443 and G189.6+03.3 (Figure \ref{fig:G189_IC443_RGB}). These star clusters are respectively NGC 2175 and NGC 2168 (also known as M35) and are indicated in Figure \ref{fig:G189_IC443_RGB} and in Figure \ref{fig:WISE_overplot}.

Looking at the emission of M35 above 0.7keV, it is in accordance with the presence of many hot and blue massive stars in the cluster (O,B spectral type). Since many of these stars are massive, it is likely some could have already undergone a supernova explosion. Strong winds and supernova explosions from massive stars could have easily wiped out the dust, which is instead present around NGC 2175. A similar idea was proposed for the Galactic starburst cluster, Westerlund 1 \citep{Muno2006,Clark2008,Negueruela2010}. We extracted the spectrum of the diffuse emission around M35, masking the point sources, which seems quite well (CSTAT/DOF=1.07) described by a non equilibrium shocked plasma (VPSHOCK model in XSPEC) with temperature kT=$0.15\pm 0.1$ keV and N$_{H}$=$0.89^{+0.09}_{-0.10}$ cm$^{-2}$. In the fit, we left all the abundances free to vary, retrieving most of them as subsolar, except for Ne (Ne/Ne$_{\odot}=1.3^{+1.5}_{-0.7}$). The low value for the ionization timescale ($\tau =0.17^{+0.16}_{-0.08}$ cm$^{-3}$ s) points towards a plasma out of equilibrium condition (see Section \ref{sec:spectral analysis} for a discussion about this parameter). This is most likely due to collisionally shocked plasma, essentially due to the strong winds coming from massive stars. However, it should be also considered the fact that is plasma is constantly illuminated by the same stars, so photoionisation effects should be important. Recently, \cite{Harer2023} discussed the importance of shocks in Westerlund 1 and in general for young star clusters in the context of TeV emission associated to Galactic cosmic ray acceleration \citep[see also][]{Vieu2023}. The presence of shocked gas in our spectra of M35 is in accordance with the turbulent environment needed to accelerate particles to TeV energies.

Moving to NGC 2175, a very faint diffuse emission can be observed in X-rays, which is found spatially coincident with a strong emitting HII region visible in infrared (Figure \ref{fig:WISE_overplot}). We tried to extract the spectrum from the region indicated in Figure \ref{fig:G189_IC443_RGB}, but we found out it indistinguishable from the background. Additional observations are needed to provide a higher statistics to characterize the gas.
\section{Discussion \label{sec:Discussion}}

In the previous Section we found indications for the existence of a ubiquitous 0.7 keV plasma component which is present in all the regions analyzed. We find the column density is close to $4.0\cdot10^{21}$ cm$^{-2}$ in region 'A', 'B' and 'D' (only for the double VPSHOCK scenario). Some differences can arise between the two models in each region, resulting in slightly different absorption values, but the overall picture is a uniform absorber covering all the regions. For a detailed discussion for each region, see Section \ref{sec:spectral analysis}.

From these findings, we argue the ubiquitous emission at 0.7 keV plasma is associated to G189.6+03.3 which results to be a foreground object, as first proposed by \cite{Asaoka1994}, placed in front of IC443. In this context, the higher column density measured in region 'C' when using a recombination model, together with high expansion velocity, might indicate IC443 is a background object which is emerging from below G189.6+03.3. In addition, we clarified this aspect, adding another thermal component to the original model and finding again the 0.7 keV. This eventually shows that this component was not fitted with the previous model, probably because it is too dim, but it is there. This supports the idea that G189.6+03.3 is in front of IC443.

In \cite{Ustamujic2021} it is described how the actual shape of IC443 might be the result of the interaction with two molecular clouds. Specifically, Figure 3 of \cite{Ustamujic2021} nicely fit the scenario of IC443 emerging from below an absorber. Moreover, from recent optical/UV data, \cite{Ritchey2020} propose the star HD 254755 may be absorbed by material in foreground, possibly associated to G189.6+03.3. However, while \cite{Asaoka1994} initially proposed only a part of G189.6+03.3 overlapped with IC443, our dataset suggests this may not be true, especially since we detect an almost uniform column density. Our spectral analysis suggests instead that G189.6+03.3 is present in all the regions analyzed, including those associated to IC443 (C,D). In Figure \ref{fig:G189_IC443_RGB}, we draw a red circle surrounding G189.6+03.3 to suggest this hypothesis (its center is shown as a red cross in Figure \ref{fig:G189_IC443_zoom}). Moreover, \cite{Greco2018} recently showed that part of our region 'D' is a shock ejecta and its shape can be correlated with the direction of the proper motion of the fast moving neutron star in the South: they conclude this structure belongs to IC443 and the plasma is overionized. The same authors propose the structure can be associated to a jet feature, as we indicated in Figure \ref{fig:G189_IC443_zoom}. In the same Figure, we also observe that in the eROSITA image G189.6+03.3 stretches from West to East. It could be possible that two jet activities took place. The direction of the first is indicated as yellow line in Figure \ref{fig:G189_IC443_zoom} which interestingly cross the putative position of the unresolved source inside G189.6+03.3: this jet should be associated to the progenitor of G189.6+03.3 with its W part not visible due the very intense emission of IC443. The second jet structure is the one investigated by \cite{Greco2018} and should be related to IC443.

\begin{figure*}
    \centering
    \includegraphics[width=7cm]{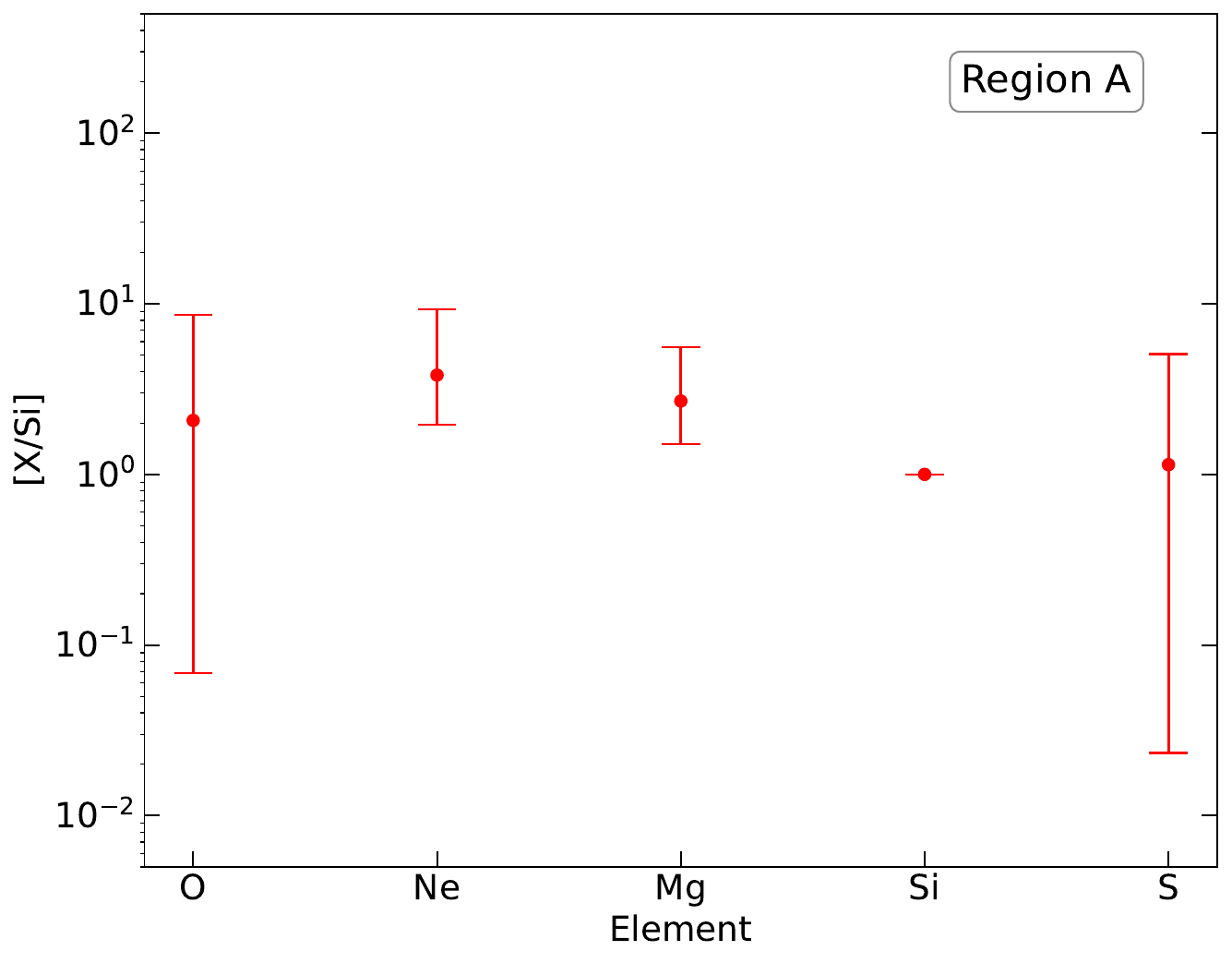}
    \includegraphics[width=7cm]{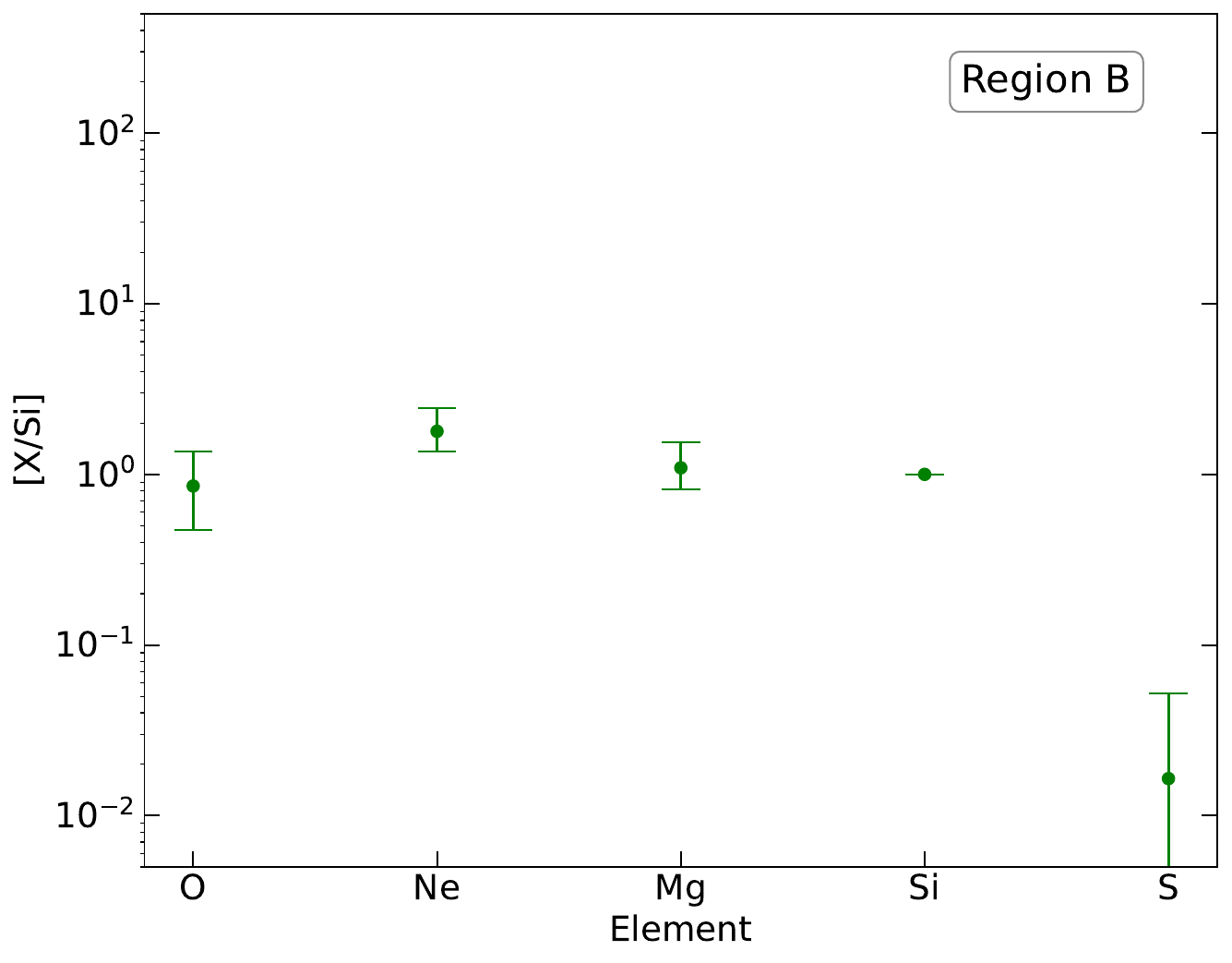}
    \includegraphics[width=7cm]{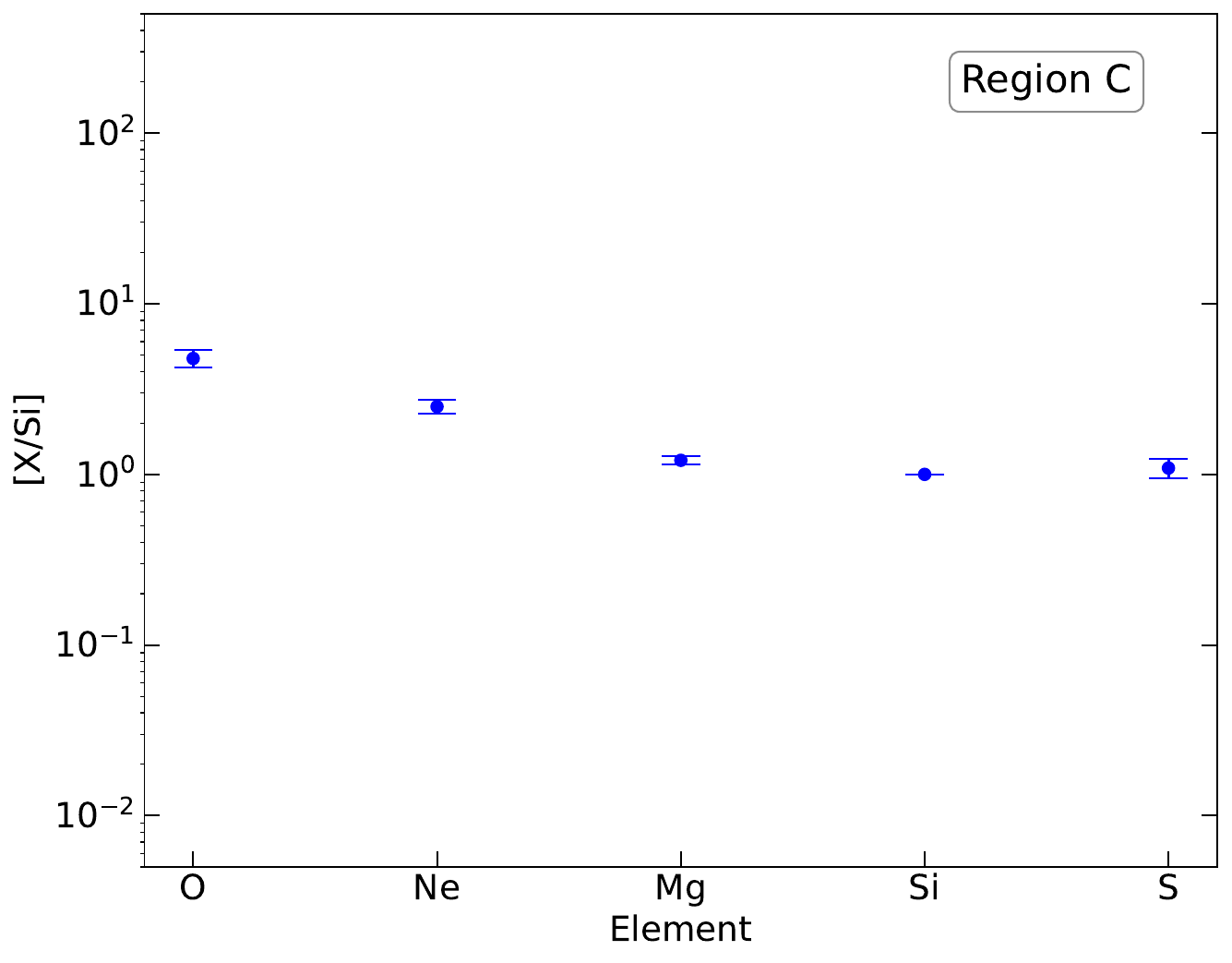}
    \includegraphics[width=7cm]{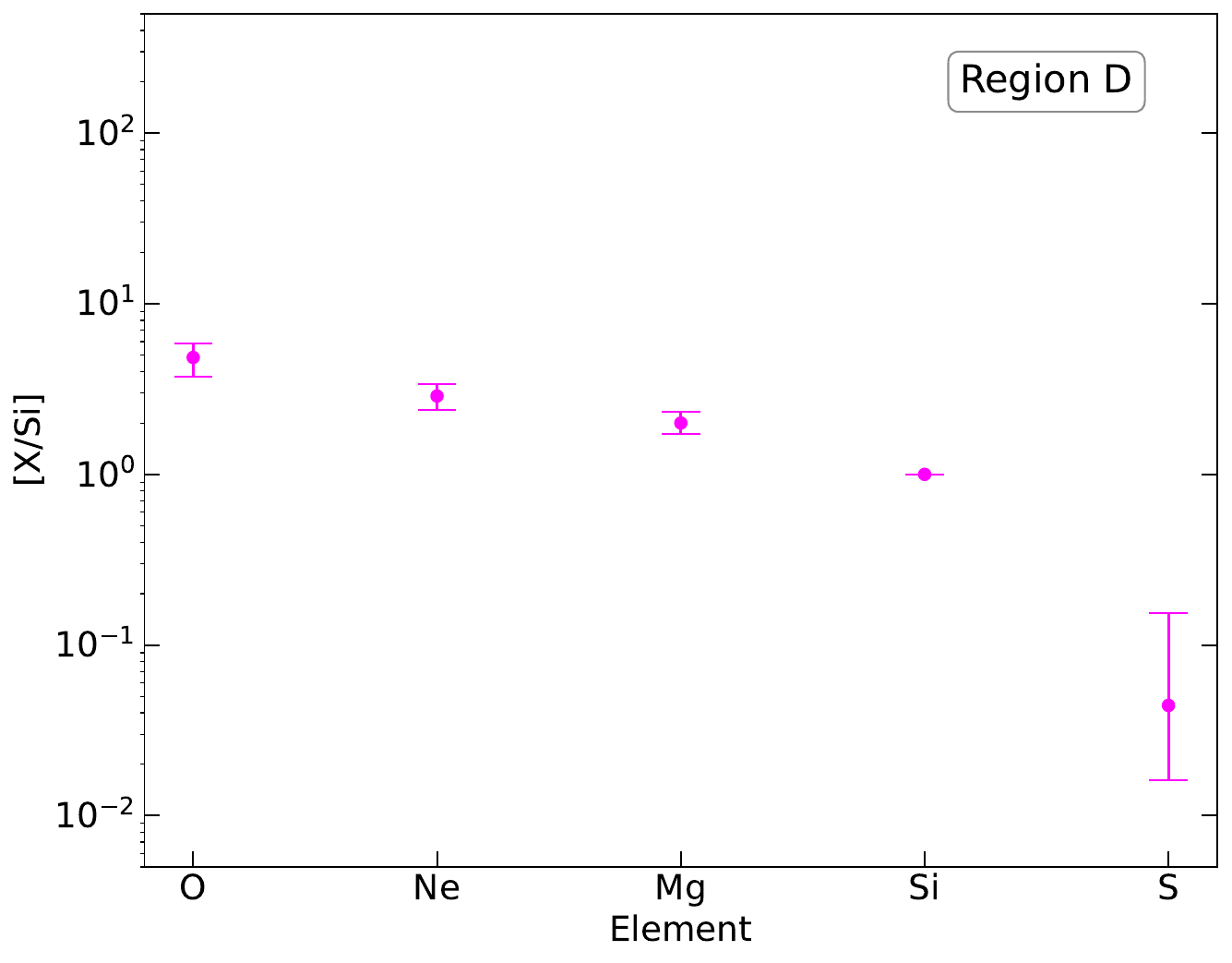}

    \caption{Logarithmic ratio of the abundance of the element X (O, Ne, Mg, Si, S) with Si. The abundances are derived from a two component fit employing two shock models (VPSHOCK).  
    \label{fig:Abundance_2vpshock}}
\end{figure*}

\begin{figure*}
    \centering
    \includegraphics[width=7cm]{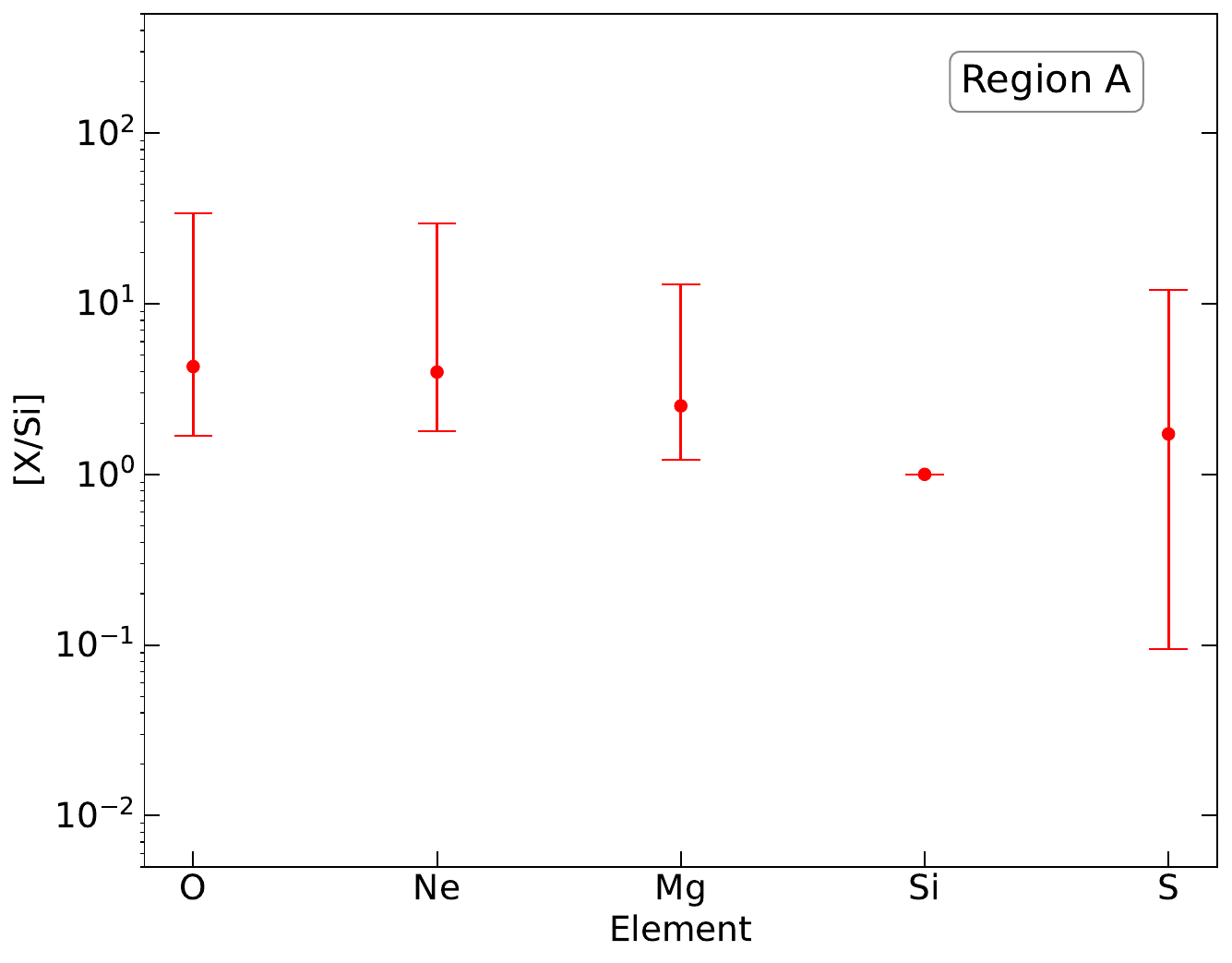}
    \includegraphics[width=7cm]{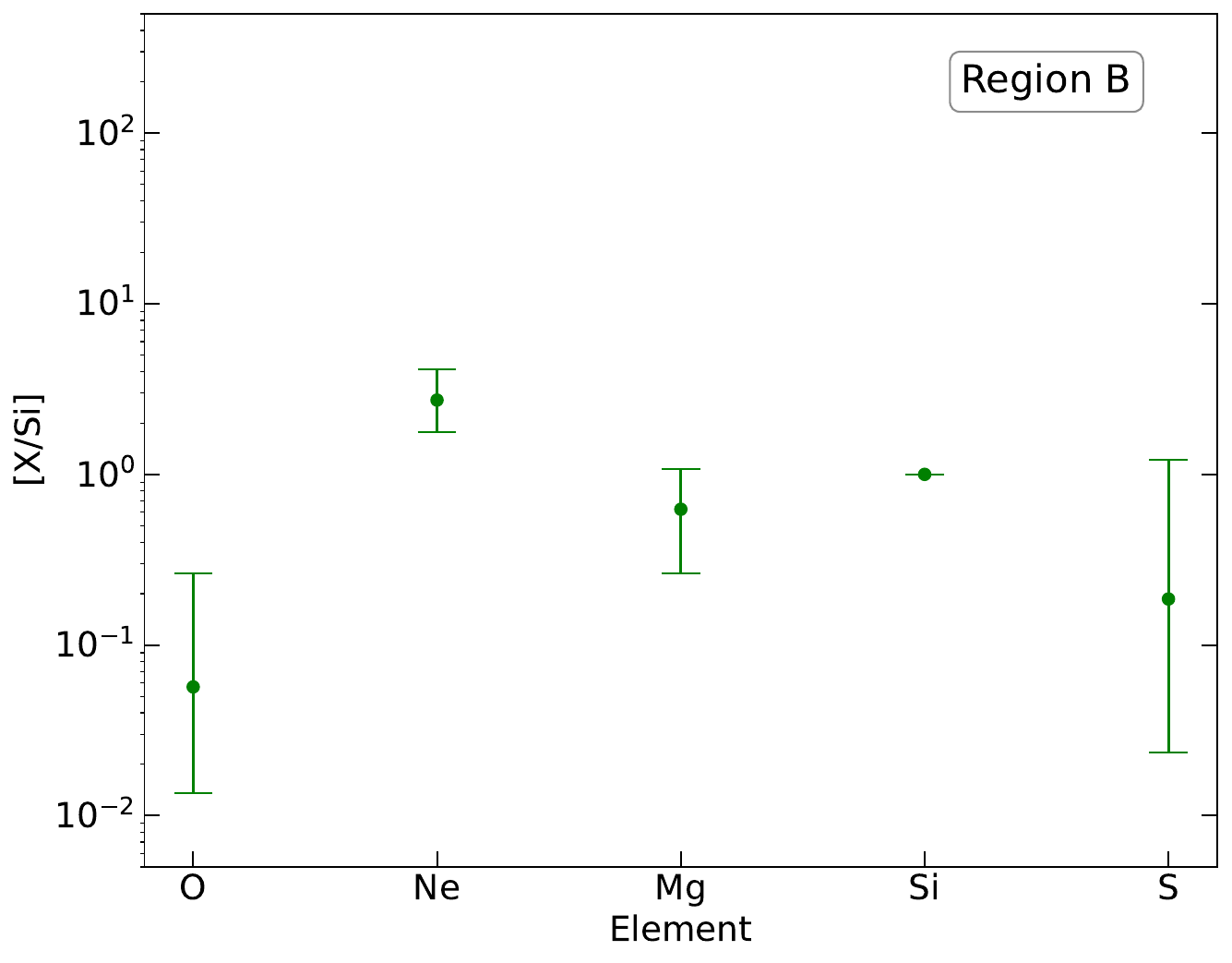}
    \includegraphics[width=7cm]{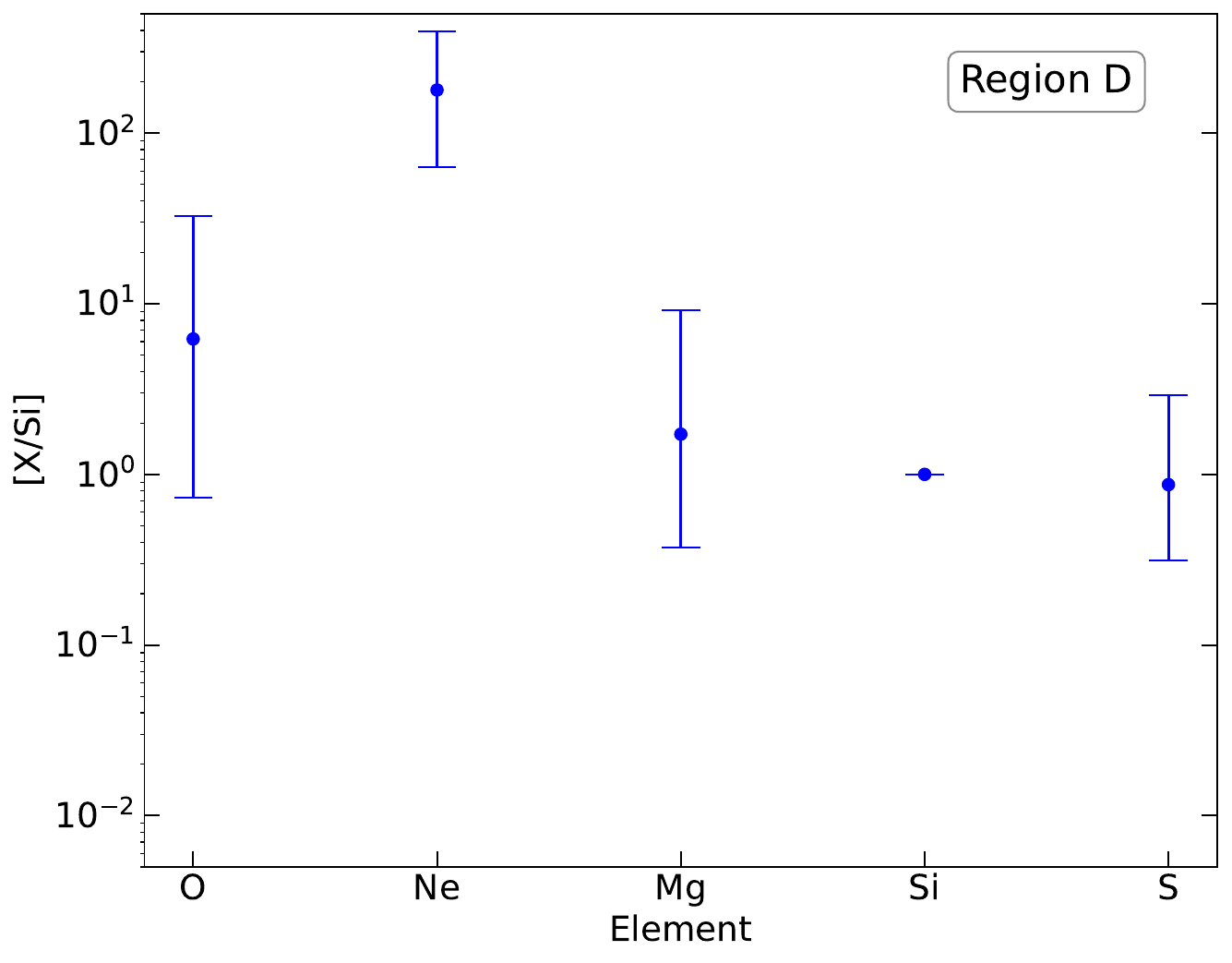}
    
    \caption{Logarithmic ratio of the abundance of the element X (O, Ne, Mg, Si, S) with Si. The abundances are derived from a two component fit employing one single temperature parallel shock model (VPSHOCK) and a recombination additive model (VRNEI). We do not show the plot of region C since the 0.7 keV components is absent when VRNEI is employed\label{fig:Abundance_vrnei}}. 
\end{figure*}

Starting from this interesting jet scenario, we want to figure out the possible types of progenitors. According to \cite{Smartt2009}, the progenitors having jets are very massive stars with M$>30$ M$_{\odot}$, specifically Luminous Blue Variable (LBV) stars. However, recent papers like \cite{Chiotellis2021} and \cite{Ustamujic2021LBV} demonstrated the influence of massive progenitors in shaping the Circumstellar Medium (CSM) through the action of winds. The final effect can be an elongated shape similar to what is expected to be created by a jet and enriched ejecta. As first observational fact, we observe supersolar abundances for O, Ne, Mg and Si in the 0.7 keV plasma component. These abundances are close to what is described in \cite{Ustamujic2021LBV} for a Luminous Blue Variable case. On the contrary of what is predicted by this model, we detect subsolar iron abundances, but this can be easily explained by a poor modeling of the Fe L-shell lines, which are not resolved in eROSITA. Moreover, the effective area of eROSITA strongly decreases above 2 keV making almost unsuitable to observe the strong Fe lines expected to arise from iron rich ejecta. Nevertheless, the faint supernova explosion model presents many of the features we highlighted above. Specifically, such kind of supernovae are predicted to have high abundance ratios in the range [C/Fe]-[Al/Fe] as a consequence of a high amount of fallback material \citep{Nomoto2013} and a jet-like structure. We observe indeed an elongated structure stretching from SE to NW in Figure \ref{fig:G189_IC443_zoom}. We therefore evaluated the ratio between the abundance of O, Ne, Mg, Si and S with Si per each region. Ideally, we would have employed Fe but since it is almost not resolved in our data, we decided to employ Si which is the element produced immediately before Fe during the explosion of a massive star. We show the results in Figure \ref{fig:Abundance_2vpshock} and in Figure \ref{fig:Abundance_vrnei}. We observe that for O, Ne and Mg the ratio is above 1 in several regions analyzed, as predicted with Fe in faint supernovae, for both the models tested.

Nevertheless, so far we have considered only single star explosion models from core collapse supernovae, but most of the stars are in binary systems. This is especially true for massive stars, which are likely to be born in crowded gas rich areas of the Galaxy. Therefore, supernova explosions driven by binary interaction can be a very common channel to explode stars \citep[see for a recent discussion][]{Laplace2021}. In Section \ref{sec:spectral analysis}, we showed how the enhancement of the temperature of the plasma in region B might indicate G189.6+03.3 is interacting with the HII region S249, especially looking at Figure \ref{fig:WISE_overplot}. Since it is well known that also IC443 is interacting with this region \citep{Fesen1984,Ambrocio2017}, this suggests the two remnants are interacting with the same HII region, i.e. they are at the same distance, regardless the progenitor was in a single or binary system. This would also be consistent with observing similar optical extinction measurement values in different points of the region, as described in Section \ref{sec:Spatial Analysis}. To test this scenario, one can assume a common distance (d=1500 pc) for the two remnants and estimate the velocity of a hypothetical compact object associated to G189.6+03.3. As an explosion site, we consider the center of the red circle (RA:06h18m37.3s, DEC:+22:14:41.3) in Figure \ref{fig:G189_IC443_RGB} and shown as red cross in Figure \ref{fig:G189_IC443_zoom}. As a current position of the object, we employ the coordinates of the diffuse emission (RA:06h19m40.8s, DEC:+21:58:03) described in Section \ref{sec:diffuse_emission_center_G189} and visible in Figure \ref{fig:G189_IC443_zoom}. Assuming 30 kyr as time of explosion, the resulting velocity is around 315km/s. This value could be tested in the future with dedicated pointed observation on the bright spot. Assuming the same explosion site for J061705.3+22212, the fast moving neutron star located close to the center of IC443, but assuming 3kyr as age and the today position, we obtain a velocity of 3200 km/s. Considering the typical proper motion values found for other compact objects by \cite{Mayer2021} are on average much lower than 3000 km/s, we argue that some different mechanism than a simple natal kick from the supernova explosion should be in place. The mechanism we propose to justify this very high proper motion value is a slingshot effect. Given the evidence provided for the existence of two separate remnants, it is appealing to consider a condition for a progenitor hosted in a system with three or more stars. Different works \cite[see for example][]{Thompson2011,Naoz2016,Hamers2022} have shown how the trajectories of the stars in such systems can be severely altered by the presence of a third star. The effect is even more chaotic when more stars are considered. Therefore, a slingshot effect might be a suitable the explanation for finding one progenitor star very far from the other and a compact object having a very high proper motion value.

The slingshot mechanism would also explain why the proper motion of J061705.3+2221 is not aligned with the center of IC443 \citep{Swartz2015,Greco2018} nor with the one of G189.6+03.3. Moreover, as shown by \cite{Ustamujic2021}, the center of the emission of IC443 determined by the maximum intensity of the X-ray emission is probably the result of a complex interplay between the expanding shock wave and a molecular cloud. Therefore, it is likely that the direction of the compact object is not aligned with it. Having two supernova explosions in two different regions would explain why the column density in region C is different comparing to the values measured in the other regions. However, it is possible also that two single stars belonging to the same association of stars, Gem OB1, originated IC443 and G189.6+03.3 respectively. In this respect, the simulations of \cite{Ustamujic2021} show how also the single explosion scenario can explain the observed shape of IC443.

We want also to discuss how several papers \citep{Yamaguchi2009,Matsumura2017} have shown the presence of recombining plasma in the region of IC443. We also obtain good fits with such model superimposed a shocked material. Interestingly, \cite{Yamauchi2020} detect a two component recombining plasma in the North-East spot of G189.6+03.3 with \textit{Suzaku}, one of which has a temperature of 0.7 keV. It is remarkable that values close to such temperature can be found also in all regions we tested and is even more remarkable for region 'D' which is quite far from the spot observed by \cite{Yamauchi2020}. Comparing to \cite{Yamauchi2020}, in this region we find an ionization timescale of the same order of magnitude with the VPSHOCK+VRNEI model, but with a poorer fit statistics comparing to our double VPSHOCK model. To explain such difference, we notice how \cite{Tanaka2022} recently underlined the contribution that charge exchange can have in supernova remnant X-ray spectra. To highlight such effect, the authors employ high resolution spectra, which have a resolving power that our instrument cannot reach. Therefore, differences in the composition and density between the clouds in region B and D might actually result in different contribution of the charge exchange that cannot be resolved in our spectra. Since charge exchange is likely to occur at the contact point between shock front and neutral material, this might explain the difference between our fit and those presented by \cite{Yamauchi2020}. Nevertheless, considering eROSITA is more sensitive in soft X-rays and instead \textit{Suzaku} collects more photons in hard X-rays, the results can be considered quite consistent among themselves.

In the same area of the sky there are also two different compact objects which we tried to understand whether they can be associated to IC443 and G189.6+03.3. Starting from J061705.3+22212, the column density of 0.7$\cdot10^{21}$ cm$^{-2}$ reported in \cite{Greco2018} is similar to what we obtain in region 'C' (Section \ref{sec:spectral analysis}). Therefore, the most straightforward conclusion is that J061705.3+22212 is probably associated to IC443, as shown by many previous studies \citep{Keohane1997,Swartz2015,Greco2018}. Regarding the pulsar PSR B0611+22, the proper motion direction rules out completely the possibility that this object is correlated to G189.6+03.3 or to IC443. Despite that, the column density derived in Section \ref{sec:Spatial Analysis} is consistent with the column density measured for G189.6+03.3 ($\sim 4.0\cdot10^{21}$ cm$^{-2}$). This would imply that the source is at least as distant as the remnants, which do not exclude the possibility that it is located much further away, as indicated from some models of the dispersion measure. Therefore, given the proper motion direction and the great uncertainties in the dispersion measure models, it is less speculative to assume that this pulsar is not associated to any of the two supernova remnants. 

In addition to these two compact objects, \cite{Bykov2008} and \cite{Zhang2018} showed the presence of several point sources at E boundary of our region C, thanks to deep \textit{Chandra}, \textit{XMM-Newton} and \textit{Nustar} observations. The nature of these objects is still quite unclear, especially since at least two of them are found to be variable. Moreover, we observe that one of these objects (dubbed 'src1a' in \cite{Bykov2008}) looks like a neutron star embedded in a pulsar wind nebula. Further studies are needed to assess whether it can be associated to IC443 or G189.6+03.3.

\section{Conclusions \label{sec:Conclusions}}
In this work, we finally confirmed that G189.6+03.3 is a supernova remnant. We find its emission can be represented by a two thermal component plasma, one of which is represented by a 0.7 keV temperature gas in equilibrium, which is found in all the regions analyzed. If we consider also that we detect a uniform absorption over the entire remnant close to $\sim 4.0\cdot10^{21}$ cm$^{-2}$, we argue a unique diffuse emission covers the whole system. The high surface brightness of region C (Figure \ref{fig:G189_IC443_RGB}) complicated the detection of the covering 0.7 keV component, which we manage to find with an additional component APEC component to our models. Given the ubiquitous presence of this plasma at the equilibrium, we conclude G189.6+03.3 completely overlaps with IC443.

We obtain high abundance ratios of [O/Si], [Ne/Si] and [Mg/Si] in most of the regions and an elongated structure, all indications in favor of a faint supernova explosion. From observing an enhancement of the temperature in the second plasma component of region B we consider the possibility that G189.6+03.3 is interacting with the HII region S249, as it is doing IC443. In this case, the two remnants should be placed at the same distance, presenting two possibilities: in one scenario, two isolated massive stars belonging to the group Gem OB1 generated the two remnants. An alternative and intriguing scenario, is given instead by two objects belonging to a multiple system. 

Given these two hypotheses, we discuss the association to the remnants of the two nearby compact objects, confirming CXOU J061705.3+22212 can be associated to IC443 while the pulsar PSR B0611+22 is unrelated to any of the two remnants. However, we suggest a third compact object could be in the field, seen as unresolved faint emission near the center of G189.6+03.3 (Section \ref{sec:diffuse_emission_center_G189}). 

Nevertheless, we also report how several unidentified point sources were observed in the past in the East part of IC443 and how it may be also possible that one of them is a compact object associated to one of the two remnants.

We conclude, underlining the necessity of new pointed observations on G189.6+03.3, given the large amount of new features shown in this work. A pointed observation toward the center would be really striking to assess the existence of a compact object, helping to shed light on whether the progenitor was indeed a faint supernova or the explosion happened via a different channel.

\begin{acknowledgements}
 We thank the anonymous referee for the useful comments and suggestions that helped to improve the quality of the manuscript. We would like to thank all the eROSITA team for the helpful discussions and suggestions provided during the realization of the paper. The image in Figure \ref{fig:WISE_overplot} has been extracted with Aladin Desktop \citep{Aladin} tool and replotted using \texttt{astropy}. FC acknowledges support from the Deutsche Forschungsgemeinschaft through the grant BE 1649/11-1 and from the International Max-Planck Research School on Astrophysics at the Ludwig-Maximilians University (IMPRS). FC thanks Hans-Thomas Janka for the useful discussion about supernova explosion models and abundance yields. WB thanks James Turner for pointing out the observation of the source location of CXCO J061705.3+22212 in the FAST GPPS. 
 This work is based on data from eROSITA, the soft X-ray instrument aboard \textit{SRG}, a joint Russian-German science mission supported by the Russian Space Agency (Roskosmos), in the interests of the Russian Academy of Sciences represented by its Space Research Institute (IKI), and the Deutsches Zentrum für Luft- und Raumfahrt (DLR). The \textit{SRG} spacecraft was built by Lavochkin Association (NPOL) and its subcontractors, and is operated by NPOL with support from the Max Planck Institute for Extraterrestrial Physics (MPE). The development and construction of the eROSITA X-ray instrument was led by MPE, with contributions from the Dr. Karl Remeis Observatory Bamberg \& ECAP (FAU Erlangen-Nuernberg), the University of Hamburg Observatory, the Leibniz Institute for Astrophysics Potsdam (AIP) and the Institute for Astronomy and Astrophysics of the University of Tübingen, with the support of DLR and the Max Planck Society. The Argelander Institute for Astronomy of the University of Bonn and the
Ludwig Maximilians Universität Munich also participated in the science preparation for eROSITA. The eROSITA data shown here were processed using the eSASS software system developed by the German eROSITA consortium. This work makes use of the Astropy Python package\footnote{\url{https://www.astropy.org/}} \citep{Astropy,Astropy2018}. A particular mention goes to the in-development coordinated package of Astropy for region handling called Regions\footnote{\url{https://github.com/astropy/regions}}. We acknowledge also the use of Python packages Matplotlib \citep{Matplotlib}, PyLaTex\footnote{\url{https://github.com/JelteF/PyLaTeX/}} and NumPy \citep{Numpy}. 
\end{acknowledgements}
\bibliography{bibliography.bib}

\begin{thebibliography}{}
\expandafter\ifx\csname natexlab\endcsname\relax\def\natexlab#1{#1}\fi
\providecommand{\url}[1]{\href{#1}{#1}}
\providecommand{\dodoi}[1]{doi:~\href{http://doi.org/#1}{\nolinkurl{#1}}}
\providecommand{\doeprint}[1]{\href{http://ascl.net/#1}{\nolinkurl{http://ascl.net/#1}}}
\providecommand{\doarXiv}[1]{\href{https://arxiv.org/abs/#1}{\nolinkurl{https://arxiv.org/abs/#1}}}

\bibitem[{{Ambrocio-Cruz} {et~al.}(2017){Ambrocio-Cruz}, {Rosado}, {de la Fuente}, {Silva}, \& {Blanco-Pi{\~n}on}}]{Ambrocio2017}
{Ambrocio-Cruz}, P., {Rosado}, M., {de la Fuente}, E., {Silva}, R., \& {Blanco-Pi{\~n}on}, A. 2017, \mnras, 472, 51, \dodoi{10.1093/mnras/stx1936}

\bibitem[{{Arnaud}(1996)}]{Xspec}
{Arnaud}, K.~A. 1996, in Astronomical Society of the Pacific Conference Series, Vol. 101, Astronomical Data Analysis Software and Systems V, ed. G.~H. {Jacoby} \& J.~{Barnes}, 17

\bibitem[{{Asaoka} \& {Aschenbach}(1994)}]{Asaoka1994}
{Asaoka}, I., \& {Aschenbach}, B. 1994, \aap, 284, 573

\bibitem[{{Astropy Collaboration} {et~al.}(2013){Astropy Collaboration}, {Robitaille}, {Tollerud}, {Greenfield}, {Droettboom}, {Bray}, {Aldcroft}, {Davis}, {Ginsburg}, {Price-Whelan}, {Kerzendorf}, {Conley}, {Crighton}, {Barbary}, {Muna}, {Ferguson}, {Grollier}, {Parikh}, {Nair}, {Unther}, {Deil}, {Woillez}, {Conseil}, {Kramer}, {Turner}, {Singer}, {Fox}, {Weaver}, {Zabalza}, {Edwards}, {Azalee Bostroem}, {Burke}, {Casey}, {Crawford}, {Dencheva}, {Ely}, {Jenness}, {Labrie}, {Lim}, {Pierfederici}, {Pontzen}, {Ptak}, {Refsdal}, {Servillat}, \& {Streicher}}]{Astropy}
{Astropy Collaboration}, {Robitaille}, T.~P., {Tollerud}, E.~J., {et~al.} 2013, \aap, 558, A33, \dodoi{10.1051/0004-6361/201322068}

\bibitem[{{Astropy Collaboration} {et~al.}(2018){Astropy Collaboration}, {Price-Whelan}, {Sip{\H{o}}cz}, {G{\"u}nther}, {Lim}, {Crawford}, {Conseil}, {Shupe}, {Craig}, {Dencheva}, {Ginsburg}, {VanderPlas}, {Bradley}, {P{\'e}rez-Su{\'a}rez}, {de Val-Borro}, {Aldcroft}, {Cruz}, {Robitaille}, {Tollerud}, {Ardelean}, {Babej}, {Bach}, {Bachetti}, {Bakanov}, {Bamford}, {Barentsen}, {Barmby}, {Baumbach}, {Berry}, {Biscani}, {Boquien}, {Bostroem}, {Bouma}, {Brammer}, {Bray}, {Breytenbach}, {Buddelmeijer}, {Burke}, {Calderone}, {Cano Rodr{\'\i}guez}, {Cara}, {Cardoso}, {Cheedella}, {Copin}, {Corrales}, {Crichton}, {D'Avella}, {Deil}, {Depagne}, {Dietrich}, {Donath}, {Droettboom}, {Earl}, {Erben}, {Fabbro}, {Ferreira}, {Finethy}, {Fox}, {Garrison}, {Gibbons}, {Goldstein}, {Gommers}, {Greco}, {Greenfield}, {Groener}, {Grollier}, {Hagen}, {Hirst}, {Homeier}, {Horton}, {Hosseinzadeh}, {Hu}, {Hunkeler}, {Ivezi{\'c}}, {Jain}, {Jenness}, {Kanarek}, {Kendrew}, {Kern}, {Kerzendorf}, {Khvalko}, {King}, {Kirkby}, {Kulkarni},
  {Kumar}, {Lee}, {Lenz}, {Littlefair}, {Ma}, {Macleod}, {Mastropietro}, {McCully}, {Montagnac}, {Morris}, {Mueller}, {Mumford}, {Muna}, {Murphy}, {Nelson}, {Nguyen}, {Ninan}, {N{\"o}the}, {Ogaz}, {Oh}, {Parejko}, {Parley}, {Pascual}, {Patil}, {Patil}, {Plunkett}, {Prochaska}, {Rastogi}, {Reddy Janga}, {Sabater}, {Sakurikar}, {Seifert}, {Sherbert}, {Sherwood-Taylor}, {Shih}, {Sick}, {Silbiger}, {Singanamalla}, {Singer}, {Sladen}, {Sooley}, {Sornarajah}, {Streicher}, {Teuben}, {Thomas}, {Tremblay}, {Turner}, {Terr{\'o}n}, {van Kerkwijk}, {de la Vega}, {Watkins}, {Weaver}, {Whitmore}, {Woillez}, {Zabalza}, \& {Astropy Contributors}}]{Astropy2018}
{Astropy Collaboration}, {Price-Whelan}, A.~M., {Sip{\H{o}}cz}, B.~M., {et~al.} 2018, \aj, 156, 123, \dodoi{10.3847/1538-3881/aabc4f}

\bibitem[{{Bocchino} \& {Bykov}(2001)}]{Bocchino2001}
{Bocchino}, F., \& {Bykov}, A.~M. 2001, \aap, 376, 248, \dodoi{10.1051/0004-6361:20010882}

\bibitem[{{Bonnarel} {et~al.}(2000){Bonnarel}, {Fernique}, {Bienaym{\'e}}, {Egret}, {Genova}, {Louys}, {Ochsenbein}, {Wenger}, \& {Bartlett}}]{Aladin}
{Bonnarel}, F., {Fernique}, P., {Bienaym{\'e}}, O., {et~al.} 2000, \aaps, 143, 33, \dodoi{10.1051/aas:2000331}

\bibitem[{{Borkowski} {et~al.}(2001){Borkowski}, {Lyerly}, \& {Reynolds}}]{Borkowski2001}
{Borkowski}, K.~J., {Lyerly}, W.~J., \& {Reynolds}, S.~P. 2001, \apj, 548, 820, \dodoi{10.1086/319011}

\bibitem[{{Borkowski} {et~al.}(1994){Borkowski}, {Sarazin}, \& {Blondin}}]{Borkowski1994}
{Borkowski}, K.~J., {Sarazin}, C.~L., \& {Blondin}, J.~M. 1994, \apj, 429, 710, \dodoi{10.1086/174355}

\bibitem[{{Braun} \& {Strom}(1986)}]{Braun1986}
{Braun}, R., \& {Strom}, R.~G. 1986, \aap, 164, 193

\bibitem[{{Brunner} {et~al.}(2022){Brunner}, {Liu}, {Lamer}, {Georgakakis}, {Merloni}, {Brusa}, {Bulbul}, {Dennerl}, {Friedrich}, {Liu}, {Maitra}, {Nandra}, {Ramos-Ceja}, {Sanders}, {Stewart}, {Boller}, {Buchner}, {Clerc}, {Comparat}, {Dwelly}, {Eckert}, {Finoguenov}, {Freyberg}, {Ghirardini}, {Gueguen}, {Haberl}, {Kreykenbohm}, {Krumpe}, {Osterhage}, {Pacaud}, {Predehl}, {Reiprich}, {Robrade}, {Salvato}, {Santangelo}, {Schrabback}, {Schwope}, \& {Wilms}}]{Brunner2022}
{Brunner}, H., {Liu}, T., {Lamer}, G., {et~al.} 2022, \aap, 661, A1, \dodoi{10.1051/0004-6361/202141266}

\bibitem[{{Burton} {et~al.}(1988){Burton}, {Geballe}, {Brand}, \& {Webster}}]{Burton1988}
{Burton}, M.~G., {Geballe}, T.~R., {Brand}, P.~W.~J.~L., \& {Webster}, A.~S. 1988, \mnras, 231, 617, \dodoi{10.1093/mnras/231.3.617}

\bibitem[{{Bykov} {et~al.}(2008){Bykov}, {Krassilchtchikov}, {Uvarov}, {Bloemen}, {Bocchino}, {Dubner}, {Giacani}, \& {Pavlov}}]{Bykov2008}
{Bykov}, A.~M., {Krassilchtchikov}, A.~M., {Uvarov}, Y.~A., {et~al.} 2008, \apj, 676, 1050, \dodoi{10.1086/529117}

\bibitem[{{Cash}(1979)}]{Cash1979}
{Cash}, W. 1979, \apj, 228, 939, \dodoi{10.1086/156922}

\bibitem[{{Chiotellis} {et~al.}(2021){Chiotellis}, {Boumis}, \& {Spetsieri}}]{Chiotellis2021}
{Chiotellis}, A., {Boumis}, P., \& {Spetsieri}, Z.~T. 2021, \mnras, 502, 176, \dodoi{10.1093/mnras/staa3573}

\bibitem[{{Clark} {et~al.}(2008){Clark}, {Muno}, {Negueruela}, {Dougherty}, {Crowther}, {Goodwin}, \& {de Grijs}}]{Clark2008}
{Clark}, J.~S., {Muno}, M.~P., {Negueruela}, I., {et~al.} 2008, \aap, 477, 147, \dodoi{10.1051/0004-6361:20077186}

\bibitem[{{Claussen} {et~al.}(1997){Claussen}, {Frail}, {Goss}, \& {Gaume}}]{Claussen1997}
{Claussen}, M.~J., {Frail}, D.~A., {Goss}, W.~M., \& {Gaume}, R.~A. 1997, \apj, 489, 143, \dodoi{10.1086/304784}

\bibitem[{{Cornett} {et~al.}(1977){Cornett}, {Chin}, \& {Knapp}}]{Cornett1977}
{Cornett}, R.~H., {Chin}, G., \& {Knapp}, G.~R. 1977, \aap, 54, 889

\bibitem[{{Davies} {et~al.}(1972){Davies}, {Lyne}, \& {Seiradakis}}]{Davies1972}
{Davies}, J.~G., {Lyne}, A.~G., \& {Seiradakis}, J.~H. 1972, \nat, 240, 229, \dodoi{10.1038/240229a0}

\bibitem[{{Denoyer}(1978)}]{Denoyer1978}
{Denoyer}, L.~K. 1978, \mnras, 183, 187, \dodoi{10.1093/mnras/183.2.187}

\bibitem[{{Ebeling} {et~al.}(2006){Ebeling}, {White}, \& {Rangarajan}}]{Ebeling2006}
{Ebeling}, H., {White}, D.~A., \& {Rangarajan}, F.~V.~N. 2006, \mnras, 368, 65, \dodoi{10.1111/j.1365-2966.2006.10135.x}

\bibitem[{{Fesen}(1984)}]{Fesen1984}
{Fesen}, R.~A. 1984, \apj, 281, 658, \dodoi{10.1086/162142}

\bibitem[{{Fesen} \& {Kirshner}(1980)}]{Fesen1980}
{Fesen}, R.~A., \& {Kirshner}, R.~P. 1980, \apj, 242, 1023, \dodoi{10.1086/158534}

\bibitem[{{Foreman-Mackey} {et~al.}(2013){Foreman-Mackey}, {Hogg}, {Lang}, \& {Goodman}}]{emcee2013}
{Foreman-Mackey}, D., {Hogg}, D.~W., {Lang}, D., \& {Goodman}, J. 2013, \pasp, 125, 306, \dodoi{10.1086/670067}

\bibitem[{{Foster} {et~al.}(2017){Foster}, {Smith}, \& {Brickhouse}}]{Foster2017}
{Foster}, A.~R., {Smith}, R.~K., \& {Brickhouse}, N.~S. 2017, in American Institute of Physics Conference Series, Vol. 1811, Atomic Processes in Plasmas (APiP 2016), 190005, \dodoi{10.1063/1.4975748}

\bibitem[{{Gaensler} {et~al.}(2006){Gaensler}, {Chatterjee}, {Slane}, {van der Swaluw}, {Camilo}, \& {Hughes}}]{Gaensler2006}
{Gaensler}, B.~M., {Chatterjee}, S., {Slane}, P.~O., {et~al.} 2006, \apj, 648, 1037, \dodoi{10.1086/506246}

\bibitem[{{Greco} {et~al.}(2018){Greco}, {Miceli}, {Orlando}, {Peres}, {Troja}, \& {Bocchino}}]{Greco2018}
{Greco}, E., {Miceli}, M., {Orlando}, S., {et~al.} 2018, \aap, 615, A157, \dodoi{10.1051/0004-6361/201832733}

\bibitem[{{Grichener} \& {Soker}(2017)}]{Grichener2017}
{Grichener}, A., \& {Soker}, N. 2017, \mnras, 468, 1226, \dodoi{10.1093/mnras/stx534}

\bibitem[{{Hamers} {et~al.}(2022){Hamers}, {Glanz}, \& {Neunteufel}}]{Hamers2022}
{Hamers}, A.~S., {Glanz}, H., \& {Neunteufel}, P. 2022, \apjs, 259, 25, \dodoi{10.3847/1538-4365/ac49e7}

\bibitem[{{Han} {et~al.}(2021){Han}, {Wang}, {Wang}, {Wang}, {Zhou}, {Sun}, {Yan}, {Su}, {Jing}, {Chen}, {Gao}, {Hou}, {Xu}, {Lee}, {Wang}, {Jiang}, {Xu}, {Yan}, {Gan}, {Guan}, {Huang}, {Jiang}, {Li}, {Men}, {Sun}, {Wang}, {Wang}, {Wang}, {Xie}, {Xu}, {Yao}, {You}, {Yu}, {Yuan}, {Yuen}, {Zhang}, \& {Zhu}}]{Han2021}
{Han}, J.~L., {Wang}, C., {Wang}, P.~F., {et~al.} 2021, Research in Astronomy and Astrophysics, 21, 107, \dodoi{10.1088/1674-4527/21/5/107}

\bibitem[{{H{\"a}rer} {et~al.}(2023){H{\"a}rer}, {Reville}, {Hinton}, {Mohrmann}, \& {Vieu}}]{Harer2023}
{H{\"a}rer}, L.~K., {Reville}, B., {Hinton}, J., {Mohrmann}, L., \& {Vieu}, T. 2023, \aap, 671, A4, \dodoi{10.1051/0004-6361/202245444}

\bibitem[{Harris {et~al.}(2020)Harris, Millman, van~der Walt, Gommers, Virtanen, Cournapeau, Wieser, Taylor, Berg, Smith, Kern, Picus, Hoyer, van Kerkwijk, Brett, Haldane, del R{\'{i}}o, Wiebe, Peterson, G{\'{e}}rard-Marchant, Sheppard, Reddy, Weckesser, Abbasi, Gohlke, \& Oliphant}]{Numpy}
Harris, C.~R., Millman, K.~J., van~der Walt, S.~J., {et~al.} 2020, Nature, 585, 357, \dodoi{10.1038/s41586-020-2649-2}

\bibitem[{{He} {et~al.}(2013){He}, {Ng}, \& {Kaspi}}]{He2013}
{He}, C., {Ng}, C.~Y., \& {Kaspi}, V.~M. 2013, \apj, 768, 64, \dodoi{10.1088/0004-637X/768/1/64}

\bibitem[{{Humphreys}(1978)}]{Humphreys1978}
{Humphreys}, R.~M. 1978, \apjs, 38, 309, \dodoi{10.1086/190559}

\bibitem[{Hunter(2007)}]{Matplotlib}
Hunter, J.~D. 2007, Computing in Science \& Engineering, 9, 90, \dodoi{10.1109/MCSE.2007.55}

\bibitem[{{Kaastra}(2017)}]{Kaastra2017}
{Kaastra}, J.~S. 2017, \aap, 605, A51, \dodoi{10.1051/0004-6361/201629319}

\bibitem[{{Keohane} {et~al.}(1997){Keohane}, {Petre}, {Gotthelf}, {Ozaki}, \& {Koyama}}]{Keohane1997}
{Keohane}, J.~W., {Petre}, R., {Gotthelf}, E.~V., {Ozaki}, M., \& {Koyama}, K. 1997, \apj, 484, 350, \dodoi{10.1086/304329}

\bibitem[{{Lallement} {et~al.}(2019){Lallement}, {Babusiaux}, {Vergely}, {Katz}, {Arenou}, {Valette}, {Hottier}, \& {Capitanio}}]{Lallement2019}
{Lallement}, R., {Babusiaux}, C., {Vergely}, J.~L., {et~al.} 2019, \aap, 625, A135, \dodoi{10.1051/0004-6361/201834695}

\bibitem[{{Laplace} {et~al.}(2021){Laplace}, {Justham}, {Renzo}, {G{\"o}tberg}, {Farmer}, {Vartanyan}, \& {de Mink}}]{Laplace2021}
{Laplace}, E., {Justham}, S., {Renzo}, M., {et~al.} 2021, \aap, 656, A58, \dodoi{10.1051/0004-6361/202140506}

\bibitem[{{Leahy}(2004)}]{Leahy2004}
{Leahy}, D.~A. 2004, \aj, 127, 2277, \dodoi{10.1086/382241}

\bibitem[{{Manchester} {et~al.}(2005){Manchester}, {Hobbs}, {Teoh}, \& {Hobbs}}]{Manchester2005}
{Manchester}, R.~N., {Hobbs}, G.~B., {Teoh}, A., \& {Hobbs}, M. 2005, \aj, 129, 1993, \dodoi{10.1086/428488}

\bibitem[{{Matsumura} {et~al.}(2017){Matsumura}, {Tanaka}, {Uchida}, {Okon}, \& {Tsuru}}]{Matsumura2017}
{Matsumura}, H., {Tanaka}, T., {Uchida}, H., {Okon}, H., \& {Tsuru}, T.~G. 2017, \apj, 851, 73, \dodoi{10.3847/1538-4357/aa9bdf}

\bibitem[{{Mayer} \& {Becker}(2021)}]{Mayer2021}
{Mayer}, M. G.~F., \& {Becker}, W. 2021, \aap, 651, A40, \dodoi{10.1051/0004-6361/202141119}

\bibitem[{{Medan} {et~al.}(2021){Medan}, {L{\'e}pine}, \& {Hartman}}]{Medan2021}
{Medan}, I., {L{\'e}pine}, S., \& {Hartman}, Z. 2021, \aj, 161, 234, \dodoi{10.3847/1538-3881/abe878}

\bibitem[{{Micelotta} {et~al.}(2016){Micelotta}, {Dwek}, \& {Slavin}}]{Micelotta2016}
{Micelotta}, E.~R., {Dwek}, E., \& {Slavin}, J.~D. 2016, \aap, 590, A65, \dodoi{10.1051/0004-6361/201527350}

\bibitem[{{Mitsuda} {et~al.}(2007){Mitsuda}, {Bautz}, {Inoue}, {Kelley}, {Koyama}, {Kunieda}, {Makishima}, {Ogawara}, {Petre}, {Takahashi}, {Tsunemi}, {White}, {Anabuki}, {Angelini}, {Arnaud}, {Awaki}, {Bamba}, {Boyce}, {Brown}, {Chan}, {Cottam}, {Dotani}, {Doty}, {Ebisawa}, {Ezoe}, {Fabian}, {Figueroa}, {Fujimoto}, {Fukazawa}, {Furusho}, {Furuzawa}, {Gendreau}, {Griffiths}, {Haba}, {Hamaguchi}, {Harrus}, {Hasinger}, {Hatsukade}, {Hayashida}, {Henry}, {Hiraga}, {Holt}, {Hornschemeier}, {Hughes}, {Hwang}, {Ishida}, {Ishisaki}, {Isobe}, {Itoh}, {Iyomoto}, {Kahn}, {Kamae}, {Katagiri}, {Kataoka}, {Katayama}, {Kawai}, {Kilbourne}, {Kinugasa}, {Kissel}, {Kitamoto}, {Kohama}, {Kohmura}, {Kokubun}, {Kotani}, {Kotoku}, {Kubota}, {Madejski}, {Maeda}, {Makino}, {Markowitz}, {Matsumoto}, {Matsumoto}, {Matsuoka}, {Matsushita}, {McCammon}, {Mihara}, {Misaki}, {Miyata}, {Mizuno}, {Mori}, {Mori}, {Morii}, {Moseley}, {Mukai}, {Murakami}, {Murakami}, {Mushotzky}, {Nagase}, {Namiki}, {Negoro}, {Nakazawa}, {Nousek}, {Okajima},
  {Ogasaka}, {Ohashi}, {Oshima}, {Ota}, {Ozaki}, {Ozawa}, {Parmar}, {Pence}, {Porter}, {Reeves}, {Ricker}, {Sakurai}, {Sanders}, {Senda}, {Serlemitsos}, {Shibata}, {Soong}, {Smith}, {Suzuki}, {Szymkowiak}, {Takahashi}, {Tamagawa}, {Tamura}, {Tamura}, {Tanaka}, {Tashiro}, {Tawara}, {Terada}, {Terashima}, {Tomida}, {Torii}, {Tsuboi}, {Tsujimoto}, {Tsuru}, {Turner}, {Ueda}, {Ueno}, {Ueno}, {Uno}, {Urata}, {Watanabe}, {Yamamoto}, {Yamaoka}, {Yamasaki}, {Yamashita}, {Yamauchi}, {Yamauchi}, {Yaqoob}, {Yonetoku}, \& {Yoshida}}]{Mitsuda2007}
{Mitsuda}, K., {Bautz}, M., {Inoue}, H., {et~al.} 2007, \pasj, 59, S1, \dodoi{10.1093/pasj/59.sp1.S1}

\bibitem[{{Muno} {et~al.}(2006){Muno}, {Law}, {Clark}, {Dougherty}, {de Grijs}, {Portegies Zwart}, \& {Yusef-Zadeh}}]{Muno2006}
{Muno}, M.~P., {Law}, C., {Clark}, J.~S., {et~al.} 2006, \apj, 650, 203, \dodoi{10.1086/507175}

\bibitem[{{Naoz}(2016)}]{Naoz2016}
{Naoz}, S. 2016, \araa, 54, 441, \dodoi{10.1146/annurev-astro-081915-023315}

\bibitem[{{Negueruela} {et~al.}(2010){Negueruela}, {Clark}, \& {Ritchie}}]{Negueruela2010}
{Negueruela}, I., {Clark}, J.~S., \& {Ritchie}, B.~W. 2010, \aap, 516, A78, \dodoi{10.1051/0004-6361/201014032}

\bibitem[{{Nomoto} {et~al.}(2013){Nomoto}, {Kobayashi}, \& {Tominaga}}]{Nomoto2013}
{Nomoto}, K., {Kobayashi}, C., \& {Tominaga}, N. 2013, \araa, 51, 457, \dodoi{10.1146/annurev-astro-082812-140956}

\bibitem[{{Okon} {et~al.}(2021){Okon}, {Tanaka}, {Uchida}, {Tsuru}, {Seta}, {Kokusho}, \& {Smith}}]{Okon2021}
{Okon}, H., {Tanaka}, T., {Uchida}, H., {et~al.} 2021, \apj, 921, 99, \dodoi{10.3847/1538-4357/ac1e2c}

\bibitem[{{Olbert} {et~al.}(2001){Olbert}, {Clearfield}, {Williams}, {Keohane}, \& {Frail}}]{Olbert2001}
{Olbert}, C.~M., {Clearfield}, C.~R., {Williams}, N.~E., {Keohane}, J.~W., \& {Frail}, D.~A. 2001, \apjl, 554, L205, \dodoi{10.1086/321708}

\bibitem[{{Predehl} \& {Schmitt}(1995)}]{Predehl1995}
{Predehl}, P., \& {Schmitt}, J.~H.~M.~M. 1995, \aap, 500, 459

\bibitem[{{Predehl} {et~al.}(2021){Predehl}, {Andritschke}, {Arefiev}, {Babyshkin}, {Batanov}, {Becker}, {B{\"o}hringer}, {Bogomolov}, {Boller}, {Borm}, {Bornemann}, {Br{\"a}uninger}, {Br{\"u}ggen}, {Brunner}, {Brusa}, {Bulbul}, {Buntov}, {Burwitz}, {Burkert}, {Clerc}, {Churazov}, {Coutinho}, {Dauser}, {Dennerl}, {Doroshenko}, {Eder}, {Emberger}, {Eraerds}, {Finoguenov}, {Freyberg}, {Friedrich}, {Friedrich}, {F{\"u}rmetz}, {Georgakakis}, {Gilfanov}, {Granato}, {Grossberger}, {Gueguen}, {Gureev}, {Haberl}, {H{\"a}lker}, {Hartner}, {Hasinger}, {Huber}, {Ji}, {Kienlin}, {Kink}, {Korotkov}, {Kreykenbohm}, {Lamer}, {Lomakin}, {Lapshov}, {Liu}, {Maitra}, {Meidinger}, {Menz}, {Merloni}, {Mernik}, {Mican}, {Mohr}, {M{\"u}ller}, {Nandra}, {Nazarov}, {Pacaud}, {Pavlinsky}, {Perinati}, {Pfeffermann}, {Pietschner}, {Ramos-Ceja}, {Rau}, {Reiffers}, {Reiprich}, {Robrade}, {Salvato}, {Sanders}, {Santangelo}, {Sasaki}, {Scheuerle}, {Schmid}, {Schmitt}, {Schwope}, {Shirshakov}, {Steinmetz}, {Stewart}, {Str{\"u}der},
  {Sunyaev}, {Tenzer}, {Tiedemann}, {Tr{\"u}mper}, {Voron}, {Weber}, {Wilms}, \& {Yaroshenko}}]{Predehl2021}
{Predehl}, P., {Andritschke}, R., {Arefiev}, V., {et~al.} 2021, \aap, 647, A1, \dodoi{10.1051/0004-6361/202039313}

\bibitem[{{Ritchey} {et~al.}(2020){Ritchey}, {Jenkins}, {Federman}, {Rice}, {Caprioli}, \& {Wallerstein}}]{Ritchey2020}
{Ritchey}, A.~M., {Jenkins}, E.~B., {Federman}, S.~R., {et~al.} 2020, \apj, 897, 83, \dodoi{10.3847/1538-4357/ab96ce}

\bibitem[{{Salvato} {et~al.}(2022){Salvato}, {Wolf}, {Dwelly}, {Georgakakis}, {Brusa}, {Merloni}, {Liu}, {Toba}, {Nandra}, {Lamer}, {Buchner}, {Schneider}, {Freund}, {Rau}, {Schwope}, {Nishizawa}, {Klein}, {Arcodia}, {Comparat}, {Musiimenta}, {Nagao}, {Brunner}, {Malyali}, {Finoguenov}, {Anderson}, {Shen}, {Ibarra-Medel}, {Trump}, {Brandt}, {Urry}, {Rivera}, {Krumpe}, {Urrutia}, {Miyaji}, {Ichikawa}, {Schneider}, {Fresco}, {Boller}, {Haase}, {Brownstein}, {Lane}, {Bizyaev}, \& {Nitschelm}}]{Salvato2022}
{Salvato}, M., {Wolf}, J., {Dwelly}, T., {et~al.} 2022, \aap, 661, A3, \dodoi{10.1051/0004-6361/202141631}

\bibitem[{{Smartt}(2009)}]{Smartt2009}
{Smartt}, S.~J. 2009, \araa, 47, 63, \dodoi{10.1146/annurev-astro-082708-101737}

\bibitem[{{Smith} {et~al.}(2001){Smith}, {Brickhouse}, {Liedahl}, \& {Raymond}}]{Smith2001}
{Smith}, R.~K., {Brickhouse}, N.~S., {Liedahl}, D.~A., \& {Raymond}, J.~C. 2001, \apjl, 556, L91, \dodoi{10.1086/322992}

\bibitem[{{Sunyaev} {et~al.}(2021){Sunyaev}, {Arefiev}, {Babyshkin}, {Bogomolov}, {Borisov}, {Buntov}, {Brunner}, {Burenin}, {Churazov}, {Coutinho}, {Eder}, {Eismont}, {Freyberg}, {Gilfanov}, {Gureyev}, {Hasinger}, {Khabibullin}, {Kolmykov}, {Komovkin}, {Krivonos}, {Lapshov}, {Levin}, {Lomakin}, {Lutovinov}, {Medvedev}, {Merloni}, {Mernik}, {Mikhailov}, {Molodtsov}, {Mzhelsky}, {M{\"u}ller}, {Nandra}, {Nazarov}, {Pavlinsky}, {Poghodin}, {Predehl}, {Robrade}, {Sazonov}, {Scheuerle}, {Shirshakov}, {Tkachenko}, \& {Voron}}]{Sunyaev2021}
{Sunyaev}, R., {Arefiev}, V., {Babyshkin}, V., {et~al.} 2021, \aap, 656, A132, \dodoi{10.1051/0004-6361/202141179}

\bibitem[{{Swartz} {et~al.}(2015){Swartz}, {Pavlov}, {Clarke}, {Castelletti}, {Zavlin}, {Bucciantini}, {Karovska}, {van der Horst}, {Yukita}, \& {Weisskopf}}]{Swartz2015}
{Swartz}, D.~A., {Pavlov}, G.~G., {Clarke}, T., {et~al.} 2015, \apj, 808, 84, \dodoi{10.1088/0004-637X/808/1/84}

\bibitem[{{Tanaka} {et~al.}(2022){Tanaka}, {Uchida}, {Tanaka}, {Amano}, {Koshiba}, {Go Tsuru}, {Sano}, \& {Fukui}}]{Tanaka2022}
{Tanaka}, Y., {Uchida}, H., {Tanaka}, T., {et~al.} 2022, \apj, 933, 101, \dodoi{10.3847/1538-4357/ac738f}

\bibitem[{{Thompson}(2011)}]{Thompson2011}
{Thompson}, T.~A. 2011, \apj, 741, 82, \dodoi{10.1088/0004-637X/741/2/82}

\bibitem[{{Troja} {et~al.}(2008){Troja}, {Bocchino}, {Miceli}, \& {Reale}}]{Troja2008}
{Troja}, E., {Bocchino}, F., {Miceli}, M., \& {Reale}, F. 2008, \aap, 485, 777, \dodoi{10.1051/0004-6361:20079123}

\bibitem[{{Troja} {et~al.}(2006){Troja}, {Bocchino}, \& {Reale}}]{Troja2006}
{Troja}, E., {Bocchino}, F., \& {Reale}, F. 2006, \apj, 649, 258, \dodoi{10.1086/506378}

\bibitem[{{Ustamujic} {et~al.}(2021{\natexlab{a}}){Ustamujic}, {Orlando}, {Greco}, {Miceli}, {Bocchino}, {Tutone}, \& {Peres}}]{Ustamujic2021}
{Ustamujic}, S., {Orlando}, S., {Greco}, E., {et~al.} 2021{\natexlab{a}}, \aap, 649, A14, \dodoi{10.1051/0004-6361/202039940}

\bibitem[{{Ustamujic} {et~al.}(2021{\natexlab{b}}){Ustamujic}, {Orlando}, {Miceli}, {Bocchino}, {Limongi}, {Chieffi}, {Trigilio}, {Umana}, {Bufano}, {Ingallinera}, \& {Peres}}]{Ustamujic2021LBV}
{Ustamujic}, S., {Orlando}, S., {Miceli}, M., {et~al.} 2021{\natexlab{b}}, \aap, 654, A167, \dodoi{10.1051/0004-6361/202141569}

\bibitem[{{van Dyk} {et~al.}(2001){van Dyk}, {Connors}, {Kashyap}, \& {Siemiginowska}}]{vanDyk2001}
{van Dyk}, D.~A., {Connors}, A., {Kashyap}, V.~L., \& {Siemiginowska}, A. 2001, \apj, 548, 224, \dodoi{10.1086/318656}

\bibitem[{{Vieu} \& {Reville}(2023)}]{Vieu2023}
{Vieu}, T., \& {Reville}, B. 2023, \mnras, 519, 136, \dodoi{10.1093/mnras/stac3469}

\bibitem[{{Welsh} \& {Sallmen}(2003)}]{Welsh2003}
{Welsh}, B.~Y., \& {Sallmen}, S. 2003, \aap, 408, 545, \dodoi{10.1051/0004-6361:20030908}

\bibitem[{{Wilms} {et~al.}(2000){Wilms}, {Allen}, \& {McCray}}]{Wilms2000}
{Wilms}, J., {Allen}, A., \& {McCray}, R. 2000, \apj, 542, 914, \dodoi{10.1086/317016}

\bibitem[{{Yamaguchi} {et~al.}(2009){Yamaguchi}, {Ozawa}, {Koyama}, {Masai}, {Hiraga}, {Ozaki}, \& {Yonetoku}}]{Yamaguchi2009}
{Yamaguchi}, H., {Ozawa}, M., {Koyama}, K., {et~al.} 2009, \apjl, 705, L6, \dodoi{10.1088/0004-637X/705/1/L6}

\bibitem[{{Yamauchi} {et~al.}(2020){Yamauchi}, {Oya}, {Nobukawa}, \& {Pannuti}}]{Yamauchi2020}
{Yamauchi}, S., {Oya}, M., {Nobukawa}, K.~K., \& {Pannuti}, T.~G. 2020, \pasj, 72, 81, \dodoi{10.1093/pasj/psaa070}

\bibitem[{{Yao} {et~al.}(2017){Yao}, {Manchester}, \& {Wang}}]{Yao2017}
{Yao}, J.~M., {Manchester}, R.~N., \& {Wang}, N. 2017, \apj, 835, 29, \dodoi{10.3847/1538-4357/835/1/29}

\bibitem[{{Zhang} {et~al.}(2018){Zhang}, {Tang}, {Zhang}, {Sun}, {Gotthelf}, {Zhang}, {Li}, {Cheng}, {Pasham}, {Baganoff}, {Perez}, {Hailey}, \& {Mori}}]{Zhang2018}
{Zhang}, S., {Tang}, X., {Zhang}, X., {et~al.} 2018, \apj, 859, 141, \dodoi{10.3847/1538-4357/aabe7c}

\bibitem[{{Zhong} {et~al.}(2019){Zhong}, {Li}, {Carlin}, {Chen}, {Mendez}, \& {Hou}}]{Zhong2019}
{Zhong}, J., {Li}, J., {Carlin}, J.~L., {et~al.} 2019, \apjs, 244, 8, \dodoi{10.3847/1538-4365/ab3859}

\bibitem[{{Zhu} {et~al.}(2019){Zhu}, {Slane}, {Raymond}, \& {Tian}}]{Zhu2019}
{Zhu}, H., {Slane}, P., {Raymond}, J., \& {Tian}, W.~W. 2019, \apj, 882, 135, \dodoi{10.3847/1538-4357/ab3226}

\end{thebibliography}
\bibliographystyle{aasjournal}

\end{document}